\documentclass[10pt]{IEEEtran}

\usepackage{amsmath}
\usepackage{url}
\usepackage{multirow}
\usepackage{times,url,comment,cite}
\usepackage{amssymb,amsfonts,bbm}
\usepackage{graphicx,slashbox}
\usepackage{epsfig,epsf}
\usepackage{color}
\usepackage{subfigure}

\makeatletter
\def\ps@headings{%
\def\@oddhead{\mbox{}\scriptsize\rightmark \hfil \thepage}%
\def\@evenhead{\scriptsize\thepage \hfil \leftmark\mbox{}}%
\def\@oddfoot{}%
\def\@evenfoot{}}
\makeatother

\usepackage{cite}

\definecolor{marco}{rgb}{0,0.6,0}

\newtheorem{definition}{Definition}

\begin{document}


\title{Content replication and placement in mobile networks}


\author{Chi-Anh La$^\dag$, Pietro Michiardi$^\dag$, Claudio
  Casetti$^\ddag$, Carla-Fabiana Chiasserini$^\ddag$, Marco
  Fiore$^\star$\vspace*{5pt}\\
$^\dag$ EURECOM Institut, France\\
$^\ddag$ Dip. di Elettronica, Politecnico di Torino, Italy\\
$^\star$ Universit\'e de Lyon, INRIA, INSA-Lyon, CITI Lab, France\\
 Email : \{Firstname.Lastname\}@eurecom.fr 
  \\  \{Lastname\}@polito.it 
  \\  \{Firstname.Lastname\}@insa-lyon.fr 
}
\maketitle

\begin{abstract}
Performance and reliability of content access in mobile networks is 
conditioned by the number and location of content replicas deployed at the
network nodes. 
Location theory has been the traditional, \textit{centralized} approach to 
study content replication: computing the number and placement of replicas 
in a \textit{static} network can be cast as a facility location problem.
The endeavor of this work is to design a practical solution to the above
joint optimization problem that is suitable for mobile wireless environments.
We thus seek a replication algorithm that is \textit{lightweight},
\textit{distributed}, and \textit{reactive to network dynamics}.

We devise a solution that lets nodes (i) share
the burden of storing and providing content, so as to achieve load balancing,
and (ii) autonomously decide whether to replicate or drop the information, so as
to adapt the content availability to dynamic demands and time-varying network
topologies.
We evaluate our mechanism through simulation, by exploring a wide range 
of settings, including different node mobility models, content characteristics
and system scales. Furthermore, we compare our mechanism to state-of-the-art
approaches to content delivery in static and mobile networks.

Results show that our mechanism, which uses \textit{local measurements only}, 
is: (i) extremely precise in approximating an optimal solution to
content placement and replication; (ii) robust against network
mobility; (iii) flexible in accommodating various content access
patterns. Moreover, our scheme outperforms alternative approaches to
content dissemination both in terms of content access delay and access
congestion.

\end{abstract}




\section{Introduction}\label{sec:introduction}
Academic and industrial research in the networking field is pursuing
the idea that networks should provide access to contents, rather than
to hosts. Recently, this goal has been extended to wireless networks
as well, as witnessed by the tremendous growth of services and
applications offered to users equipped with advanced mobile terminals.

The inexorable consequence of a steady increase in data traffic
exerted by mobile devices fetching content from the Internet is a
drainage of network resources of mobile operators
\cite{nytimes,venutrebeat, chant09}.  A promising approach to solve
this problem is {\em content replication}, i.e., to create copies of
information content at user devices so as to exploit device-to-device
communication for content delivery. This approach has been shown to be
effective especially in wireless networks with medium-high node
density, where \textit{access congestion} is the main limiting factor
that determines the performance of content delivery (see, e.g.,
\cite{Derhab09} for a survey on the topic).

been extensively 
side, and delay 
replication 
mechanisms 
paths 
with the 

In this paper, we consider such a wireless network scenario and
explore the concept of content replication in a cooperative
environment, when the content demand and network topology dynamically
change in time.  In this context, nodes can fetch content from the
Internet using a cellular network, store it, and possibly serve other
users through device-to-device communication (e.g., using IEEE 802.11
or Bluetooth).  Our scenario also accommodates the possibility for
content to exhibit variegate popularity patterns, as well as to be
updated upon expiration of a validity-time tag, so as to maintain
consistency with copies stored by servers in the Internet.

according to an epidemic approach 
the content to all users, might not be 

The application scenario we target in this work introduces several
problems related to content replication. {\em Optimal replica
placement} is one of those: selecting the location that is better
suited to store content is difficult, especially when the network is
dynamic.  Another prominent issue is {\em how many content replicas}
should be made available to mobile nodes.  Clearly, decisions on the
placement and number of replicas to be deployed in the network are
tightly related problems: intuitively, the latter introduces a
feedback loop to the former as every content replication triggers a
new instance of the placement problem.

studied through the lenses of classic Location 
\cite{Mirchandani}.  Our endeavor is to build upon the theoretic works
that have flourished in the facility location theory literature, and
address the above \textit{joint problems}, with the ultimate goal of
designing a lightweight, distributed mechanism to achieve content
replication in mobile wireless networks.  Thus, our work departs from
previous approaches that either require global (or extended) knowledge
of the network \cite{arya01, Laoutaris07} or are
unpractical~\cite{Moscibroda05}. In particular, 
study realistic scenarios in 
simultaneously consumed by mobile nodes 
have capacity constraints for the amount 
other nodes.  we design a content replication scheme that requires
\textit{local measurements} only and that aims at evenly distributing
among nodes the demanding task of hosting a content replica and serve
others. We show that optimality in both placement and replication can
be approximated through our simple practical solution.

The contributions of this paper are summarized as follows:
\begin{itemize}
\item we revisit traditional facility location theory in the light of
the extremely challenging settings that mobile wireless networks
introduce. Leveraging the insights provided by capacitated facility
location approaches to content replication, we propose a distributed
mechanism inspired by local search approximation algorithms. Our
solution exploits a particular formulation of a multi-commodity
capacitated facility location problem to compute an approximate
solution based on local measurements only;
\item we perform an extensive simulation study where we dissect the
properties of our distributed mechanism. As a result, we show that
content placement and replication achieved through our scheme well
approximate an optimal solution when both network and content dynamics
are considered. Furthermore, our results prove that our mechanism (i)
achieves load balancing among the network nodes, in terms of both
amount of served requests and storage capacity required at each mobile
user, and (ii) scales very well with the network size and density,
making it suitable for those scenarios in which access congestion may
appear;
\item we compare through simulation our content replication scheme
with existing mechanisms, considering the realistic case where not all
users are interested in the available information items.
\end{itemize}

The remainder of the paper is organized as follows. In
Sec.~\ref{sec:background}, we give a detailed overview of the system
model and we introduce the content replication problem, pointing at
the new problems introduced by the dynamic nature of wireless
networks. In Sec.~\ref{sec:centralized}, we revisit traditional
location theory and extend it to accommodate the constraints and
requirements of our system. Based on the insights gained from a
theoretical ground, we move on to the design of our distributed
mechanism for content replication and replica placement in
Sec.~\ref{sec:system}.  In Secs.~\ref{sec:simulation} and
\ref{sec:evaluation} we describe the simulation settings and
methodology and present a thorough discussion on the results.  We
review prior works in the domain of content dissemination in mobile
networks in Sec.~\ref{sec:related-work}, and finally draw our
conclusions in Sec.~\ref{sec:conclusion}.

\section{Network scenario and problem statement}\label{sec:background}
Here, we first detail the system model we refer to,
then we define the problem of content replication 
and placement in mobile networks.
In particular, we inherit the problem of  replication typical of the
wired Internet and we
discuss the new challenges introduced by the dynamic nature of 
wireless networks with respect to their wireline counterpart.
At last, we describe the steps we take in order to address 
content replication and placement in our setting.

\subsection{System model}

We investigate a scenario including mobile
users (i.e., mobile nodes), equipped with devices offering 
3G/4G Internet connectivity as well as
device-to-device communication capabilities (e.g., through IEEE 802.11).
Although we do not concern ourselves with  the provision of Internet access  in
ad hoc wireless networks, we remark that broadband connectivity allows new
content to be fetched and, possibly, updated. 

We denote the set of mobile nodes by ${\cal V}$, with $V=|{\cal V}|$,
and we consider that they may be interested in a set of information
items. We refer to such a set as ${\cal I}$ and to its cardinality as $I$. 
Each item $i \in {\cal I}$, of size $s(i)$, is tagged with a validity
time, and originally hosted on a server in the
Internet, which can be accessed by mobile users through the broadband
access we hinted at. 
We define as $p(i)$ the content popularity level of the generic item $i$,
i.e., the fraction of nodes interested in such an item. Thus, we
have $0 \leq p(i) \leq 1$, with $p(i)=1$ corresponding to the highest popularity
level, i.e., when all nodes in the system are interested in content $i$.

We focus on a \textit{cooperative environment} where a node $j \in {\cal V}$
wishing to access the content first tries to retrieve it from  other devices.
If its search fails,
the node downloads a fresh content replica from the Internet server and
temporarily stores it for a period of time $\tau_j$,  termed
\textit{storage time}. For simplicity of presentation, in the following we
assume $\tau_j = \tau,$  $\forall j\, \in {\cal V}$. During the storage period,
$j$ serves  the content to other nodes upon receiving a request for it and, possibly,
downloads from the Internet server a fresh copy of the content if its validity
time has expired.  We refer to the nodes hosting an information copy at a
given time instant as
\textit{replica nodes}.  We denote the set of nodes storing a copy of
item $i$ at time $t$ by ${\cal R}_i(t)$, and define 
${\cal R}(t)=\cup_{i \in {\cal I}} {\cal R}_i(t)$, with $R=|{\cal
  R}|$.  
Also, we associate to each replica node $j$ a capacity value $c_j$,
which, 
as we shall see later, relates to the capability of the node to serve content requests.

A node, which is interested in a generic information item $i$ and 
does not store any copy of it,  
issues queries for such an item at a rate $\lambda$.  
Replica nodes, which receive a query for an information item they currently
store, will reply with a message including the requested  content. 

Finally, in order to clearly define the problem we address, in the
following we model the network topology at a given time instant $t$ through a
graph $G(t)=({\cal V},{\cal E}(t))$, whose set of vertices coincides
with the set of network nodes ${\cal V}$
and the set of edges ${\cal E}(t)$ represents the set of links
existing between the network nodes at time $t$.

\subsection{Problem statement}

Both content replication and caching have received significant
attention in the literature, due to their importance in enhancing performance,
availability and reliability of content access for Web-based
applications.
The two problems, however, differ since content replication is an independent
process aimed at creating copies of a content at the network nodes,
regardless of whether they asked for it or not. Caching, instead, 
is a by-product of the content query mechanism as only nodes that retrieved
the content have the possibility to cache it~\cite{Derhab09}. 

Our claim (confirmed by simulation results) is that, in a network
scenario as the one we address in this
work, content replication is to be preferred to caching. Indeed,
caching may lead to the creation of a large number of copies in the
network, especially for highly-popular content. In medium-high dense
networks, this raises the problems of (i) large overhead
due to multiple replies to a single query, (ii)
energy depletion of a large fraction of nodes acting as content
providers, (iii) congestion in accessing the cellular network
for fresher versions of the content in order to avoid inconsistencies.
We therefore deal with content replication, that is, we
design a mechanism to determine how many replicas should be
created in the network and where, under dynamic, realistic conditions.

Traditionally, a similar problem has been studied through the lenses of classic 
Facility Location Theory \cite{Mirchandani}, by considering replicas
to be created in the network as facilities to open.
Which new problems are then introduced in our work? 

\noindent {\em i)} Content replication and placement can be cast as an
optimization problem in presence of static network
conditions. However, node mobility leads to a dynamic graph $G(t)$,
which would require the problem to be solved upon every network
topology or demand rate change.

\noindent {\em ii)} While addressing content replication, we also
target load balancing among the nodes. Even under static topology and
constant demand, solving the facility location problem does not yield
load balancing.

\noindent {\em iii)} The input to the facility location problem is the
content demand workload generated by users: both replica locations and
the number of replicas to deploy in a network depend on content
consumption patterns.  While the approach traditionally adopted is to
assume content demand to be directed to the closest facility, the
wireless nature of our system yields unpredictable propagation paths
for content requests, potentially reaching multiple facilities
(replica nodes).

\noindent {\em iv)} The traditional approach defines two separate
sets, one for facilities (replica nodes) and one for the users. In our
context, instead, any node may store an information replica as well as
request an item which it does not currently own.

As a first step to address all of the above issues, in
Sec. \ref{sec:centralized} we restrict our attention to a simplified network
setting and revisit a centralized approach for facility
location problems. Our goal is to gain sufficient insights from such
a problem formulation, as well as from solutions to it
proposed in the literature, to build a distributed approach
that closely approximates the optimal solution to the problem.
Then, in Sec. \ref{sec:system} we 
consider a dynamic scenario (i.e., mobile nodes and time-varying
demand) and seek an algorithm that only requires local knowledge and a
distributed implementation.

\section{Getting insights: A centralized approach}\label{sec:centralized}
The simplified network scenario we address here is characterized by
static nodes and constant demand; furthermore, we drop the
load balancing requirement we previously outlined and assume that
content queries are directed to the closest replica node.
For simplicity, let us fix the time instant and drop the time
dependency from our notation; also, let all users be interested in
every content $i$ ($i=1,\ldots,I$) and request it at the same
constant rate. 

Given such a scenario, we formulate our replication problem as 
a {\em capacitated} facility location problem where the set of replica
nodes $\mathcal{R}=\cup_i\mathcal{R}_i$ corresponds to the set of
facilities that are required to be opened,  nodes requesting a content 
are referred
to as clients and information items correspond to the commodities that
are available at each facility. We model the capacity of a replica
node as the number of clients that a facility can serve.
The goal is to identify the \textit{subset} of facilities that,
at a given time instant, can serve the clients so as to minimize some
global cost function while satisfying the facility capacity
constraints.

We point out that, with respect to traditional formulations of the
capacitated 
facility location problem, we need to take into account the following 
aspects. Both clients and facilities
lay on the same network graph $G = (\mathcal{V}, \mathcal{E})$. As
such, any vertex of the graph can be a client or a facility: all
vertexes that are not selected as facilities will be treated as
clients. 

In the location theory
literature, two copies of the same facility can be opened at the same
location, in order to increase the capacity of a site. Instead, in our
work a vertex of the graph can host only one copy of the same facility: 
indeed, it is
reasonable to assume that a node stores only one copy of the same
information item.

For the sake of clarity, we first
define a \textit{single-commodity} capacitated facility location
problem, where we delve into the details of local search techniques
that have been applied in the literature to solve such problems. We
then move to a \textit{multi-commodity} version of the
problem and discuss the issues related to the
capacity constraints we are required to satisfy in this case.

\subsection{The single commodity problem}


Let us consider one information item only (i.e., $I=1$). Then, we
can define the single commodity capacitated facility location problem 
as follows.


\begin{definition}
  \label{def:cfl}
  Given the  set ${\cal V}$ of nodes (which can act as both clients
and facility nodes) 
  and cost $f_j$ of opening a facility at $j \in {\cal V}$,
  select a subset of nodes as facilities, ${\cal R} \subseteq {\cal V}$, 
so as to
  minimize the joint cost $C({\cal V},f)$ of opening the facilities
  and serving the demand while ensuring that each facility $j$ can
  only serve at most $c_j$ clients:
  \begin{equation}
    \label{eq:cost}
    C({\cal V},f) = \sum_{j \in \mathcal{R}}{f_j} + \sum_{h \in {\cal V}}{d(h,m_h)}. 
  \end{equation} 
In~(\ref{eq:cost}), $m_h \in \mathcal{R}$ is the facility $j$
\textit{closest} to $h$, 
and $d(h,m_h)$ is the cost attributable to facility $m_h$ for serving
client $h$ (in the literature, this is typically modelled as a
pair-wise distance function between client and facility).
Also, the number of clients attached to facility $j\in\mathcal{R}$, i.e.,
\[u_j=|\{h \in {\cal V},~\mbox{s.t.}~m_h=j\}|,\]  
must be such that $u_j \leq c_j$.
\end{definition}

In words, the above problem amounts to finding how many 
facilities should be open, and at which nodes, so as to minimize the
average distance to access a facility from a client location, while
satisfying the capacity constraints of each opened facility. 
This problem nicely translates into our setting, where we need to
establish the number of replicas to be created for an information item 
and find the best nodes to store them so as to minimize the distance (hence
the delay) to access the information.
We also point out that  the facility location
problem in Def.~\ref{def:cfl} reduces to a $k$-median problem if the number of
facilities is given, i.e., $R=k$, and we drop the capacity
constraints. 
The solution to such a special case maps to finding the
best location for $k$ facilities to be opened. 


It is well known that, for general graphs, the above problems are
NP-hard \cite{Kariv79} and a variety of approximation algorithms have
been developed and analyzed to solve them. 
Among these algorithms, the ones based on local search
are the most versatile~\cite{arya01}.  In a general form, a local search algorithm
to solve capacitated 
facility location problems consists of an iterative procedure
in which, at every step, a
variation is applied to the current solution of the
problem. If the \textit{global} cost decreases, the variation is accepted as a new
solution to the problem. The algorithm stops when no more improvements
can be obtained. 
Three variations are possible: to {\em swap} the location of a currently opened
facility,  to {\em drop} a currently opened facility, and to
{\em add} a facility to the current solution.  Note that
the local search algorithm to the capacitated version of the
facility location problem is fairly complex: indeed, it involves
the computation of a minimum cost flow problem in order to verify the
capacity constraints \cite{arya01}. 

Such local search procedures will inspire our 
distributed mechanism described in Sec.~\ref{sec:system}, where 
we introduce 
three basic operations that iteratively, albeit
\textit{asynchronously}, yield the solution to the content replication
problem. However, there are some important remarks to make. 
The key point in our solution is 
the definition of the opening costs
$f_j$'s, which  allows us to move from a centralized to a
distributed implementation as well as provide load balancing.
Moreover, the particular operation that each
node executes to solve the replica placement problem is performed
irrespectively of the number of replicas in the network. As
such, content placement and replication are effectively de-coupled.
Finally,  in our network system  adding and swapping
are constrained operations: only vertexes that are
connected by an edge to the current vertex hosting a content replica
can be selected as possible replica
locations. Thus, our operations are \textit{local} and information
item replicas can only
move by one hop at the time in the underlying network graph.

\subsection{The Multi-commodity problem}
\label{sec:multi-item}

We now consider the more general setting in which multiple commodities
(i.e., information items)
may be available at each facility (i.e., replica node).

While the problem can be defined similarly to
Def.~\ref{def:cfl}, the cost function that we need to minimize,
formerly defined in~(\ref{eq:cost}), has to  be rewritten as follows:
\begin{equation}
\label{eq:cost_multi}
   C({\cal V},f) = \sum_{i \in {\cal I}} \sum_{j \in {\cal R}_i}f_j(i)
+ \sum_{i \in {\cal I}} \sum_{h \in {\cal V}} d(h,m_h(i)) 
\end{equation} 
where $f_j(i)$ is the cost to open a facility for commodity $i$, 
${\cal R}_i \subseteq {\cal V}$ is the subset of nodes acting as 
facilities for commodity $i$, 
$m_h(i) \in \mathcal{R}_i$ is the facility holding item $i$
that is the closest to $h$, and the number 
$u_j(i)$\footnote{Clearly, we have $u_j(i)=0$ if $j$ does not own
$i$.} 
of clients requesting
any content $i$ attached to facility $j \in {\cal R}_i$,
i.e., 
$u_j(i)=|\{h \in {\cal V}~\mbox{s.t.}~m_h(i)=j\}|$,   is such that
$\sum_{i \in {\cal I}}{u_j(i)} \leq c_j$.

In the traditional formulation of such problem,
with distinct sets of facilities and clients, a solution 
amounts to finding the location and the number of facilities
to open so that the overall client requests are satisfied.  In
our setting, however, the problem is more complex: since any vertex of the
graph $G$ can host a facility or can be a client, it is possible for a
vertex to assume both roles. Indeed, a vertex can be a replica node for
one or more information items, and, at the same time, a client requesting
information items that are not currently hosted at the vertex.

Finding approximate solutions to the multi-commodity capacitated 
facility locations  is still an open issue and little is known
concerning local search heuristics that can be effectively implemented
in practice. 
In this work, we take a simple approach that
has been also discussed in \cite{multicom}: a solution to the
multi-commodity problem is built from the union of the solutions to
individual single-commodity facility location problems. Therefore, we
transform the formulation from multi-commodity to single-commodity by
solving the above problem for each item $i$
($i=1,\ldots,I$) separately.

Then, for each item $i$, (\ref{eq:cost_multi}) becomes:
 \begin{equation}
    \label{eq:cost2}
    C({\cal V},f(i)) = \sum_{j \in \mathcal{R}_i}{f_j(i)} + 
\sum_{h \in {\cal V}}{d(h,m_h(i))}
  \end{equation} 
where $m_h(i)\in \mathcal{R}_i$ is the facility
closest to $h$ and the number of clients attached to
facility $j \in {\cal R}_i$ is such that the capacity constraints
are satisfied.

Despite the apparent simplicity of such an approach, 
how the capacity constraints are verified remains an issue to be discussed.
In our work, we adopt the two techniques presented below, where we
denote the subset of commodities hosted at $j$ by ${\cal I}_j$ and
its cardinality by $I_j$:
\begin{enumerate}
\item Each opened facility has a capacity that is allocated to each
  commodity individually. In practice, this translates into having a
  separate budget allocated to each information item that is currently
  replicated at a node in the network. Formally, the capacity constraints
  can be written as $u_j(i) \leq c_j / I_j$, $\forall i \in {\cal I}_j$,
  where we equally split the budget $c_j$ available to facility $j$ over
  all the commodities it hosts. In the following, we name such a technique
  {\em split capacity budget}.
\item We consider a facility to have a capacity that is shared among
  the commodities currently hosted by the facility. This case appears
  to be more realistic for our application scenario: each node hosting
  replicas of information items allocates a preset budget that is used
  to serve all the contents requested by other nodes. Formally, we
  define the capacity constraints for this case as follows: $\sum_{i
  \in {\cal I}_j}{u_j(i)} \leq c_j$, and we refer to such a technique
  as {\em shared capacity budget}.
\end{enumerate}

In conclusion, the approach we take in this work is to break the joint
optimization problem of the capacitated multi-commodity facility
location into a number of single-commodity location problems,
as from (\ref{eq:cost2}), for which we use the local search
techniques outlined above with the additional considerations we
made in this section concerning the capacity constraints.

To the best of our knowledge, there is no known practical,
distributed algorithm to obtain approximate solutions to the
capacitated version of the multi-commodity facility location
problem either. 
In the next section, we therefore propose a new  approach that    
only requires local knowledge, which is acquired
with simple measurements, and also provides load-balancing. 
It follows that, even in a static scenario,
our distributed algorithm does not converge to a static configuration
in which a fixed set of nodes is selected to host content
replicas. As such, the traditional methods that are used in the
literature to study the convergence properties and the locality gap of
local search algorithms cannot be directly applied, which is the main
reason for us to take an experimental perspective and validate our
work through simulations.

\section{Distributed mechanism for content replication}\label{sec:system}
We now describe our distributed replication mechanism. Armed with the
insights on the problem formulation discussed in
Sec.~\ref{sec:centralized}, our mechanism mimics a local search
procedure, by allowing replica nodes to execute one of the following
three operations on the content: (1) handover, (2) replicate or (3)
drop. For clarity of presentation, in the following we describe our
mechanism in terms of two objectives: content replication 
(Sec.~\ref{sec:content-replication}) and replica placement
(Sec.~\ref{sec:cache-placement}).  Indeed, the handover operation
amounts to solving the optimal placement of content replicas, whose
number is determined through the add and drop operations.  

For simplicity, we consider again that all users are interested in
every content $i$ ($i=1,\ldots,I$) and request it at the same constant
rate. Also, we fix the time instant and drop the time dependency from
our notation.

\subsection{Content replication}
\label{sec:content-replication}


Let us define the workload of the generic replica node $j$ for content
$i$, $w_j(i)$, as the number of requests for content $i$ served by $j$
during its storage time.  Also, recall that we introduced the value
$c_j$ as the capacity value of node $j$ and we provided a definition
that suited the simplified, static scenario described in
Sec.~\ref{sec:centralized}.  We now adapt the definition of $c_j$ to
the dynamic scenario at hand, as the reference volume of data that
replica node $j$ is willing to provide during the time it acts as a
replica node, i.e., in a storage time $\tau$.  Then, with reference to
Eq.~\ref{eq:cost}, we denote by $f_j=\sum_{i \in {\cal I}_j} f_j(i)$
the cost associated with replicas at node $j$.

Given the load balance we wish to achieve across all replica nodes and
the node capacity constraint, the total workload for replica node $j$
should equal $c_j$. Thus, we write $f_j$ as:
\begin{equation}
  \label{eq:open-cost}
  f_j= c_j - \sum_{i \in {\cal I}_j} s(i)w_j(i) 
\end{equation} 
where we recall that $s(i)$ denotes the size of content $i$.  In other
words, we let the cost associated with replica node $j$ grow with the
gap between the workload experienced by $j$ and its capacity $c_j$.

Then, during storage time $\tau$, the generic replica node $j \in
{\cal R}$ measures the number of queries that it serves, i.e.,
$w_j(i)$ $\forall i \in {\cal I}_j$.  When its storage time expires,
the replica node $j$ computes $f_j$ and takes the following decisions:
if $f_j > \epsilon$ the content is \textit{dropped}, if $f_j < -
\epsilon$ the content is \textit{replicated}, otherwise the hand-over
operation is executed (see Sec.~\ref{sec:cache-placement}). Here,
$\epsilon$ is a tolerance value to avoid replication/drop decisions in
case of small changes in the node workload.


The rationale of our mechanism is the following. If $f_j < -\epsilon$,
replica node $j$ presumes that the current number of content replicas
in the area is insufficient to guarantee the desired volume of data,
hence the node replicates the content and hands the copies over to two
of its neighbors (one each), following the placement mechanism
described below in Sec.~\ref{sec:cache-placement}. The two selected
neighbors will act as replica nodes for the subsequent storage time.
Instead, if $f_j > \epsilon$, node $j$ estimates that the workload the
current number of replicas can provide is exceeding the total demand,
thus it just drops the content copy. Finally, if the experienced
workload is (about) the same as the reference value, replica node $j$
selects one of its neighbors to which to hand over the current copy,
again according to the mechanism detailed next.

\subsection{Replica placement}
\label{sec:cache-placement}


As noted in Sec.~\ref{sec:centralized}, given the graph representing
the network topology at a fixed time instant, the placement of $R=k$
replicas can be cast as a $k$-median problem.  By applying the
approximation algorithm in \cite{arya01}, we observed that the
solution of such a problem for different instances of the topology
graph yields replica placements that are instances of a random
variable uniformly distributed over the graph.  As a consequence, in a dynamic environment our
target is to design a distributed, lightweight solution that closely
approximates a uniform distribution of the replicas over the network
nodes while ensuring load balancing among them.  To this end, we
leverage some properties of random walk and devise a mechanism, called
\textit{Random-Walk Diffusion (RWD)}, that drives the ``movement'' of
replicas over the network according to a random walk mobility model.

According to RWD, at the end of its storage time $\tau$, a replica
node $j$ randomly selects another node $l$ to store the content for
the following storage period, with probability $p_{j,l}=
\frac{1}{d_j}$ if $l$ is a neighbor of $j$, and $0$ otherwise, 
where $d_j$ is the current number of neighbors of node $j$.  In this
way, each replica performs a random walk over the network, by moving
from one node to another at each time step $\tau$.  Thus, we can apply
the result stating that in a connected, non-bipartite graph, the
probability of being at a particular node $j$ converges with time to
$d_j/(2|{\cal E}|)$ \cite{SurveyGraphs}.  In other words, if the
network topology can be modeled by a regular graph\footnote{A graph is
  regular if each of its vertices has the same number of neighbors.}
with the above characteristics, the distribution of replicas moving
according to a random walk converges to a stationary distribution,
which is uniform over the nodes.

In general, real-world networks yield non-regular graphs. However,
when $V$ nodes are uniformly deployed over the network area and have
the same radio range, the node degree likely has a binomial
distribution with parameters $(V-1)$ and $p$, with $p$ being the
probability that a link exists between any two nodes
\cite{Bollobas,Hekmat}.

For practical values of $p$ and $V$ in the scenarios under study, we
verified that the node degree distribution is indeed binomial with low
variance, i.e., all nodes have similar degree.  It follows that a
random walk provides an acceptable uniform sampling of the network
nodes, hence the replica placement distribution well approximates the
uniform distribution.

A similar result can be obtained also for clustered network
topologies, where each cluster core results to be an expander graph
\cite{Benjamini06}. In this case, a uniform replica placement over the
nodes can be achieved within each of the network clusters, thus
ensuring the desired placement in all areas where the user demand is
not negligible.

Finally, we stress that the presence of $R$ replicas in the network
corresponds to $R$ parallel random walks.  As observed in
\cite{KeqinLi10}, this reduces by almost a factor $R$ the expected
time to sample all nodes in the network, which is closely related to
the time needed to approximate the stationary distribution by a
constant factor \cite{Kahn00}.  It follows that, given a generic
initial distribution of the replicas in the network, the higher the
$R$, the more quickly the replica placement approximates a uniform
distribution.



\section{Simulation scenario}\label{sec:simulation}
We implemented our mechanism in the $ns-2$ simulator.  We consider a
wireless network with high node density, namely $3.2 \cdot
10^{-4}$~nodes/m$^2$, on a square area of $1$~km$^2$, which results in
$V=320$ and an average node degree of 9.6 neighbors. By default, nodes
move according to the stationary random waypoint
model~\cite{leboudec05} with an average node speed of 1~m/s and a mean
pause time of 100~s, a setting that is representative, for example, of
customer mobility within a mall. We also explored the performance of our mechanism in presence of outdoor pedestrian mobility.

We assume nodes to be equipped with a standard 802.11 interface, with
a 54~Mbps fixed data transmission rate and a radio transmission range
of 100~m.  As our focus is on the placement and replication of items
within the ad-hoc network, we do not simulate cellular
access. However, we account for the delay associated with the download
of information items from the cellular network, by assuming a
throughput of 384 kbps, matching that typically provided by 3G
technologies to outdoor mobile users.

The rate at which a node interested in a content generates 
queries for that item is set to $\lambda=0.01$~requests/s. 
As for the propagation of the queries in the ad hoc network,
we assume the presence of a content-location service that
nodes can access to obtain the identity of the closest content replica\footnote{%
Since query propagation is not the focus of our work, we do not further
address how such a service is maintained; for details, we refer
the reader to the vast literature on the topic, e.g.,~\cite{Friedman06}
and references therein.}.
A query for the closest replica node is then propagated using sequence numbers 
to detect and discard duplicate queries, as well as an
application-driven broadcast that optimally selects the forwarding 
nodes by leveraging the Preferred Group Broadcast
(PGB) technique~\cite{naumov06}. 
Also, a TTL is included into queries,
allowing them to travel 5 hops at most so as to prevent network flooding.
Once reached by the request, the intended destination serves it, while
other replica nodes ignore the query.

As far as the content return path is concerned, we assume that, at
each hop, the identity of the last node that relayed the query is
included in the message and recorded at the following forwarder.
Thus,  the path from the target replica node to the
query source is backtracked at the application layer without resorting to ad hoc
routing protocols, which would induce overhead or delay in the process.

Since all standard MAC-layer operations are simulated, both queries and replies may
be lost due to typical problems encountered in 802.11-based ad hoc networks (e.g.,
collisions, hidden terminals): if a query fails (i.e., no answer is received
after 2~s), a new request is issued, up to a total of 5 times\footnote{According
to extensive calibration tests, omitted due to space limitations, these
parameters provide the best results in terms of content access performance.}.

Finally, concerning the replication/drop parameters, the tolerance
value $\epsilon$ used in the replication/drop algorithm is set to
5\% of the node capacity budget, while
the storage time $\tau$ is set to 100~s.


For each experiment described in the following, results are averaged over 10
simulation runs, each lasting around 3 hours of simulated time after a warm-up period
of 500~s.

\section{Results}\label{sec:evaluation}
We present the main results of our work organized in a series of
questions. 
Furthermore, in order to benchmark the 
distributed mechanism proposed in Sec.~\ref{sec:system} against the
centralized approach discussed in Sec.~\ref{sec:centralized}, we
implement the latter as follows. Given the network time evolution, we
take a snapshot of the network topology every $\tau$~s. For
every snapshot, we solve $I$ separate single-commodity problems as
in~(\ref{eq:cost2}), under both split and shared capacity budgets. To
do so, we set $f_j(i) = c_j/I_j - u_j(i)$ and $f_j= c_j -
\sum_{i \in {\cal I}_j} u_j(i)$
in the case of split and shared capacity budget respectively, with
$u_j(i) = s(i)w_j(i)$.
As a result, load balancing is achieved under
the assumption that each content query reaches one replica node only. 

\subsection{Benchmarking the replication scheme}
\label{sec:multi-content-rep}

Here, we provide baseline results on the performance of
our replication scheme with respect to the multi commodity problem
presented in Sec.~\ref{sec:multi-item}, and discuss its fairness.

\begin{figure}[t]
\centering    
\subfigure[Number of replicas]{
	\label{fig:item-joint-vs-independent}
	\includegraphics[width=0.22\textwidth]{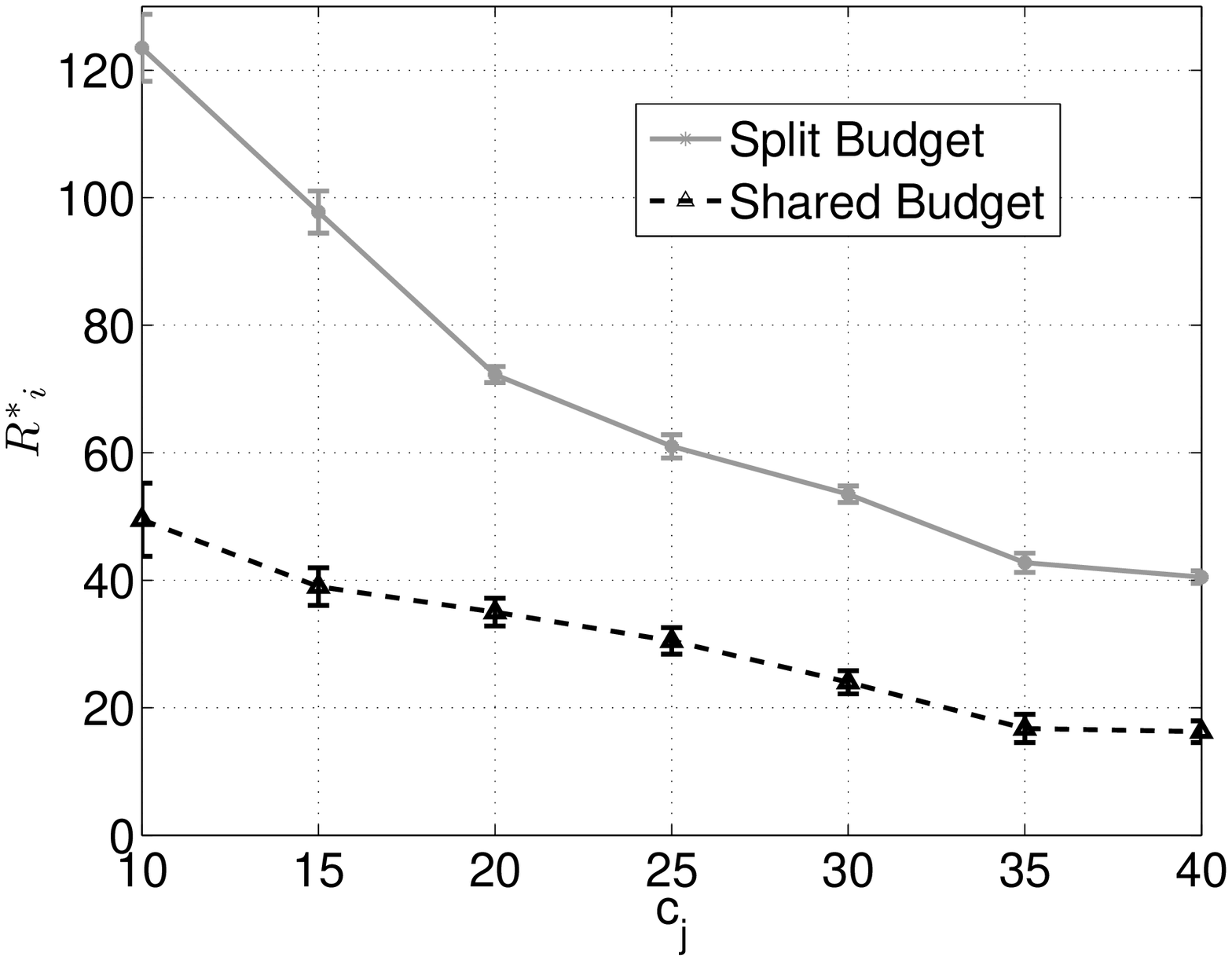}
}
\subfigure[Delay]{
	\label{fig:delay-joint-vs-independent}
	\includegraphics[width=0.22\textwidth]{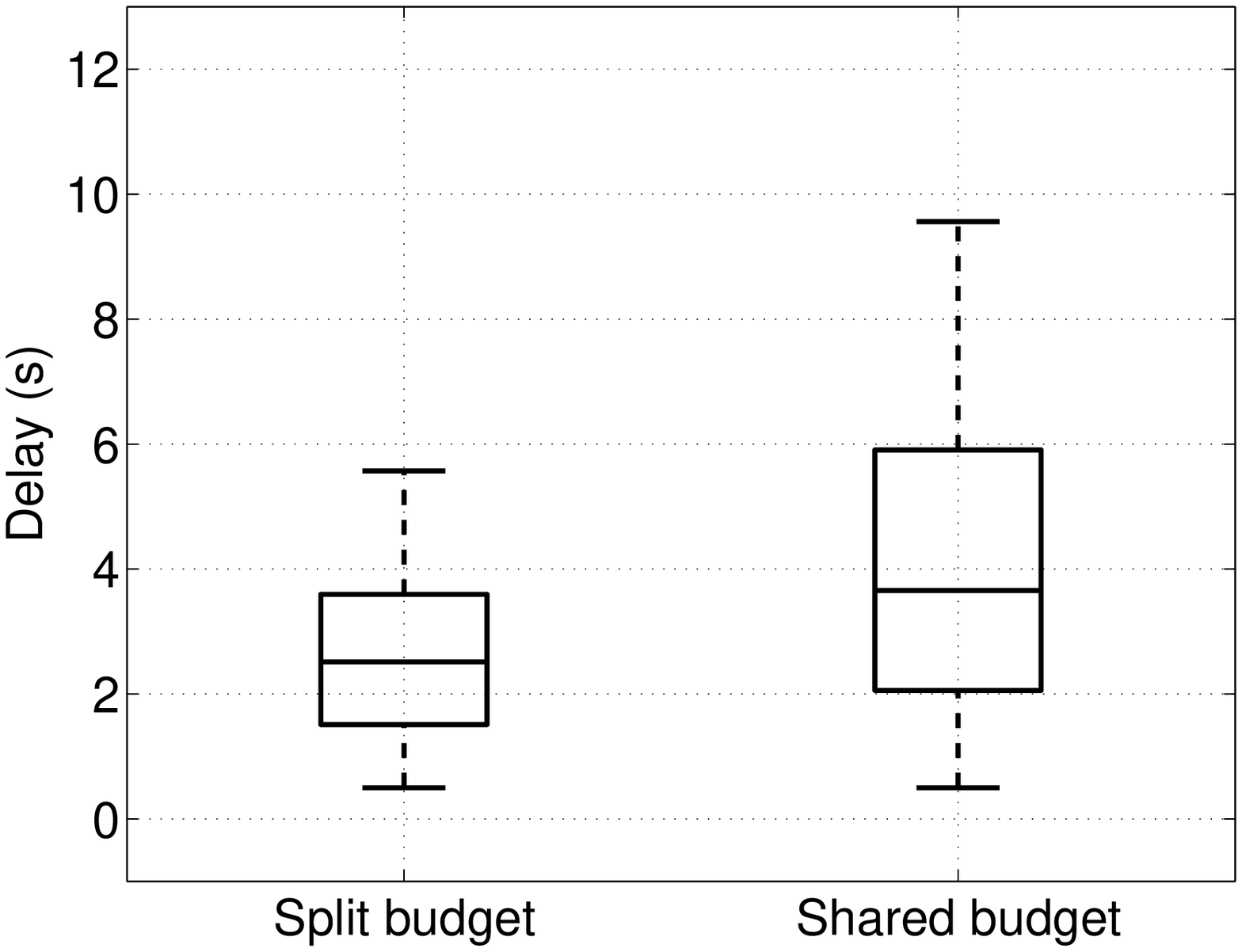}
}
\caption{Numerical solutions of the optimization problems in terms of
  number of replicas (a) and query solving delay (b)}
\label{fig:performance-joint-vs-independent}
\vspace{-3mm}
\end{figure}

\begin{figure}[t]
\subfigure[Number of replicas]{
	\label{fig:rep-joint-vs-independent}
	\includegraphics[width=0.22\textwidth]{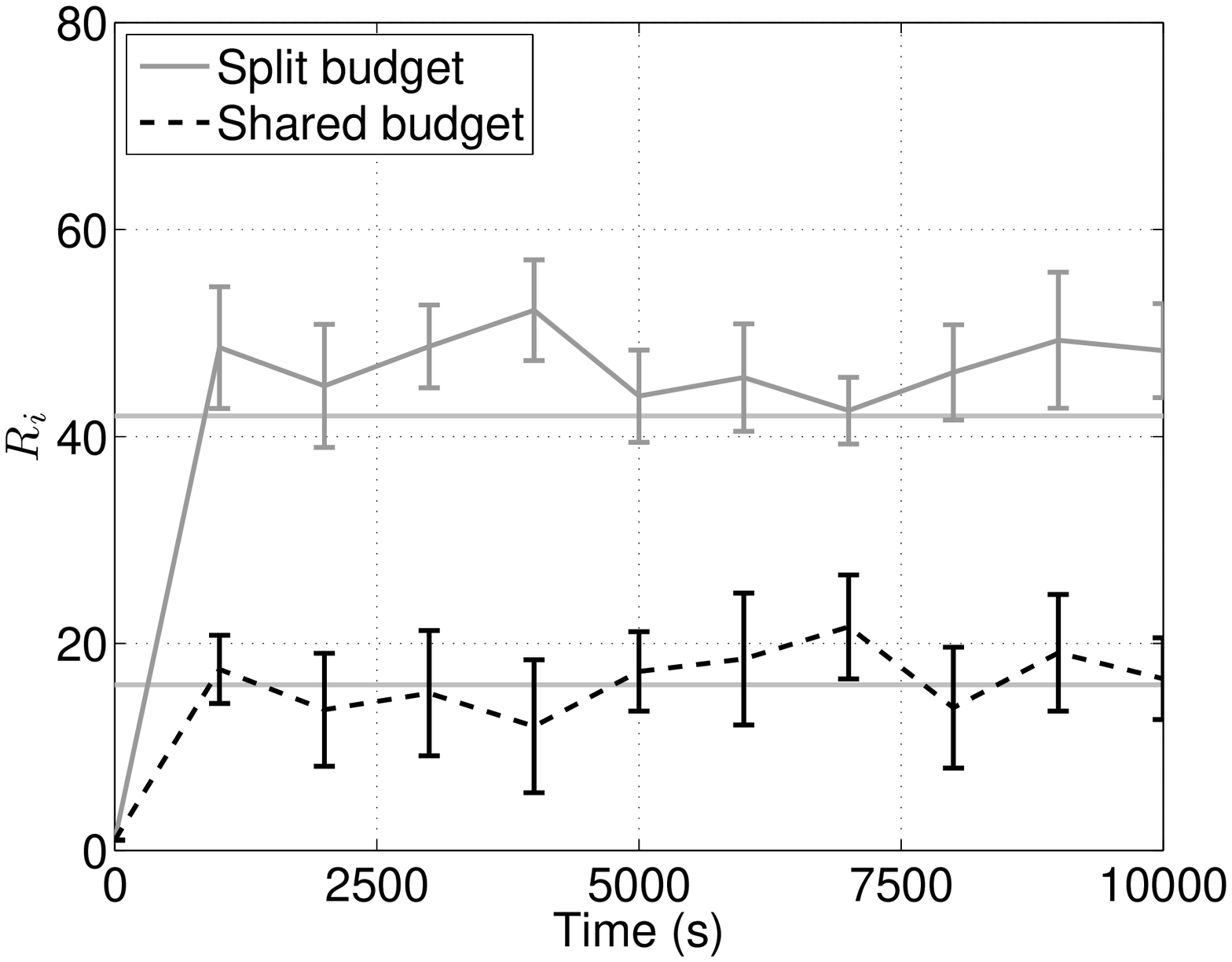}
}
\subfigure[$\chi^2$ index]{
	\label{fig:chi-joint-vs-independent}
	\includegraphics[width=0.22\textwidth]{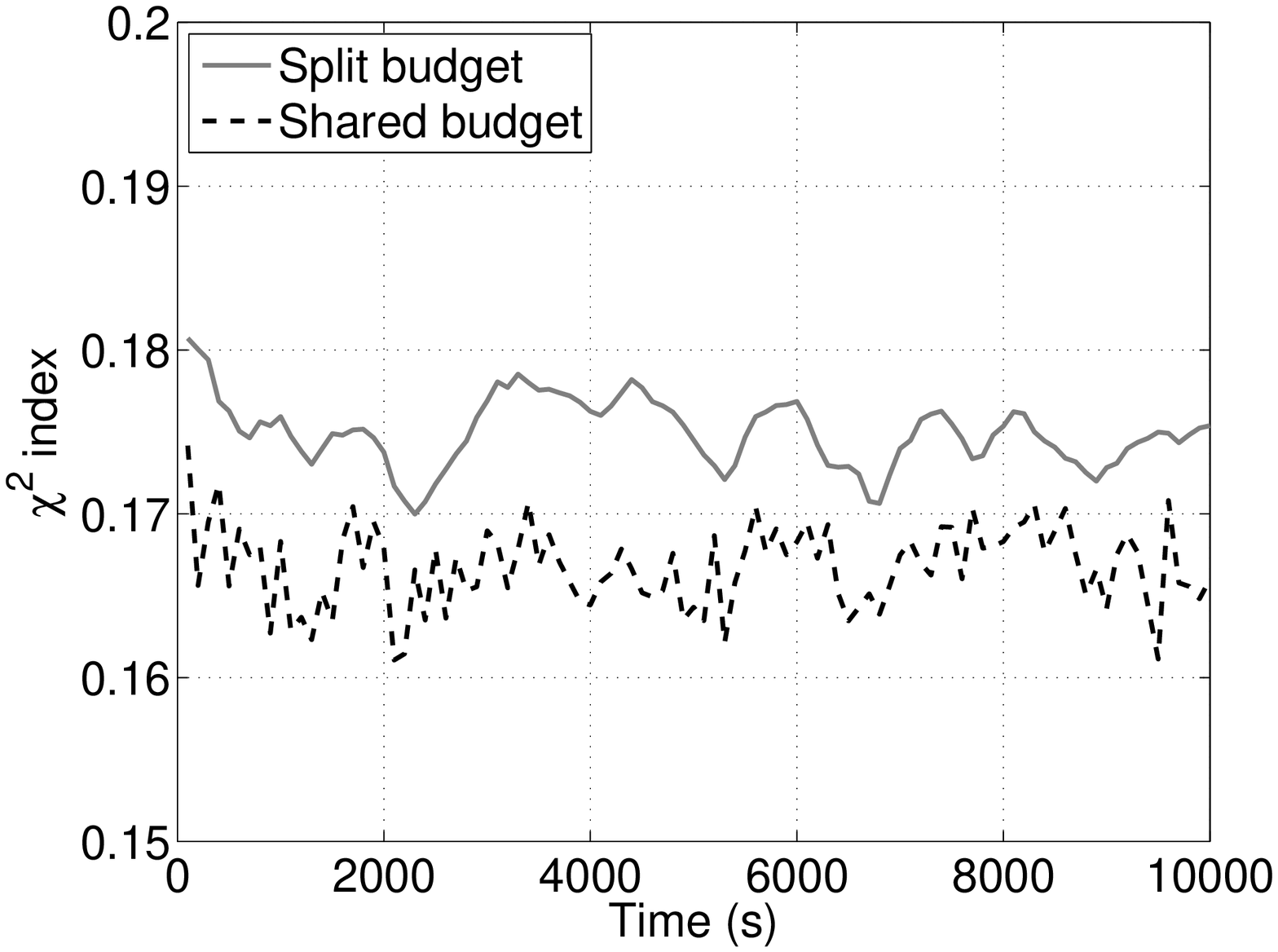}
}

\caption{Numerical solutions of the optimization problems, and comparison against our
	replication scheme: temporal evolution of the number of
        replicas (a), and of the $\chi^2$ index
        (b)}
\label{fig:performance-joint-vs-independent-vs-ours}
\vspace{-3mm}
\end{figure}

\subsubsection*{What is the impact of the capacity budget on the
  replication scheme?}
To answer this first question, we run the CFL centralized
algorithm in a snapshot of the mobile network topology, in presence
of 4 items of 1~Mbytes each. We vary the value of $c_{j}$ from 10~Mbytes to
40~Mbytes, which, in the case of optimization with split capacity budget,
means that each content is assigned a budget $c_{j}/4$.

The optimal number of replicas per information item, denoted by
$R_i^*$, is obtained by numerically solving the optimization problem
in (\ref{eq:cost2}), in both its split and shared capacity budget versions,
and is shown in Fig.~\ref{fig:item-joint-vs-independent}\footnote{Here and in
the following, unless stated otherwise, the results refer to one of
the four items since similar results were obtained for each of
them.}.
The plot clearly shows that, as higher budgets allow replica nodes to
satisfy larger amounts of requests, increasing $c_{j}$ reduces the need
for replication, with the result that a lower number of replicas is
present in the network.

It is interesting to observe that a significantly higher number of
replicas is required by an optimization with split capacity budget
with respect to that needed by an optimization with shared capacity
budget. The reason is that the latter, using a common budget for all
items, forces replications only when the total workload for all
items exceeds the budget. Conversely, optimization with split
capacity budget uses separate budgets for each content and, thus,
results in more frequent violations of such constraints.

Now, intuitively, a large number of replicas may have a beneficial effect
on content access performance: more replicas should imply higher
chances for queries to be satisfied through device-to-device
communication. 
In Fig.~\ref{fig:delay-joint-vs-independent} we show the most
important percentiles (5\%, 25\%, 50\%, 75\%, 95\%) of content access
delay with split and shared capacity constraints,
for $c_{j}=40$~Mbytes. 
Contrary to the intuition, our results indicate that the advantage
granted by a high number of replicas under the split capacity is
quite negligible, and this is mainly due to the congestion that
arises in the wireless network.

In summary, our findings pinpoint that the replication mechanism with
shared capacity constraints is a suitable approach. Beside 
experimental results, there are also practical reasons to opt for
shared capacity constraints. Indeed, in the split capacity case, a
budget has to be assigned to each item currently stored by a replica
node, which is a quantity that may vary over time. As a consequence,
content replicas may not be suitably handled if the remaining
capacity available to a node is not appropriately
re-distributed. Furthermore, usability aspects also play a role in
favor of a shared capacity approach: it would be unfeasible to ask a
user to select a service budget to allocate to every possible
item she will ever replicate.

\subsubsection*{How does our replication scheme perform with respect
to the CFL centralized algorithm?}
In order to provide an answer, we simulate our replication scheme and
we focus on the case where $c_{j}=40$~Mbytes.
As shown in Fig.~\ref{fig:rep-joint-vs-independent}, our replication scheme
can well approximate the results obtained by solving the optimization problems:
indeed, the number of replicas $R_i$ generated by our scheme
is very close to the optimal value $R_i^*$, in both the cases of split and
shared capacity budget.
Moreover, the number of replicas in the system appears quite stable
over time, which is obviously a desirable feature.

Not only the number, but also the placement of replicas itself is
important when comparing our scheme against a centralized
solution. 
Thus, we now investigate the similarity between the replica placement
achieved by our technique and that obtained with the CFL centralized
algorithm over the different snapshots representing the network
evolution. To do so, we employ the well-known  $\chi^2$ 
goodness-of-fit test on the inter-distance between content replicas\footnote{Note that 
using inter-distances instead of actual coordinates allows us to
handle a much larger  
number of samples (e.g., $V\cdot(V-1)$ instead of just $V$ samples)
thus making the  computation of the $\chi^2$ index more accurate.}. 
As depicted in
Fig.~\ref{fig:chi-joint-vs-independent}, the $\chi^2$ error obtained comparing
the distributions we achieve with the optimal ones is extremely low in
all cases; indeed, the $\chi^2$ error we obtain is well below the value
\footnote{With 14 degrees of freedom as in our case,
such value is 23.685.}
needed to accept the null hypetesis that the two
distributions are the same at a 95\% confidence level.

\subsubsection*{How fair is our replication scheme?}
\begin{figure}[t]
\centering
\subfigure[Stored items]{
	\label{fig:cdf_item-joint-vs-independent}
	\includegraphics[width=0.225\textwidth]{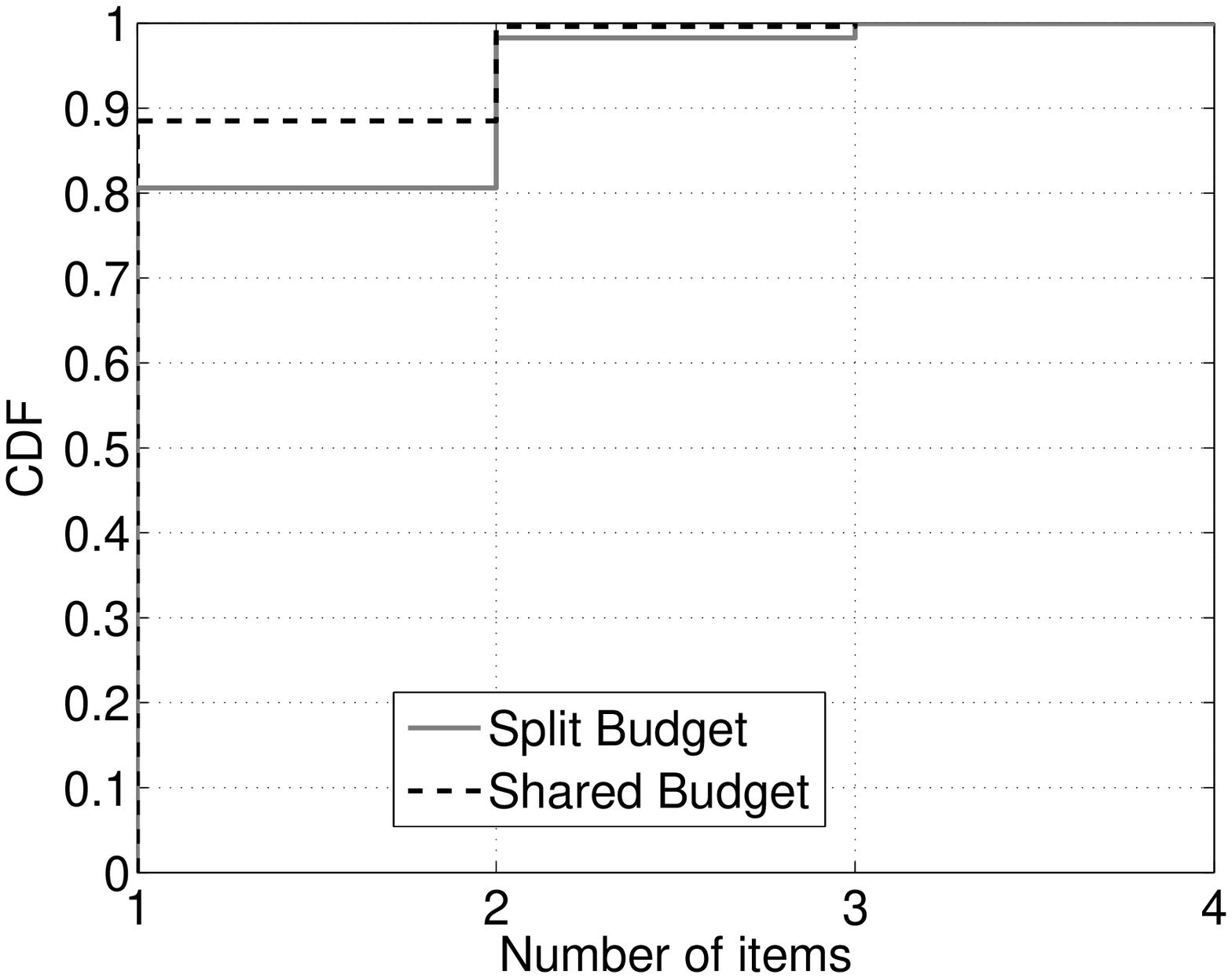}
}
\subfigure[Workload]{
	\label{fig:cdf_load-joint-vs-independent}
	\includegraphics[width=0.225\textwidth]{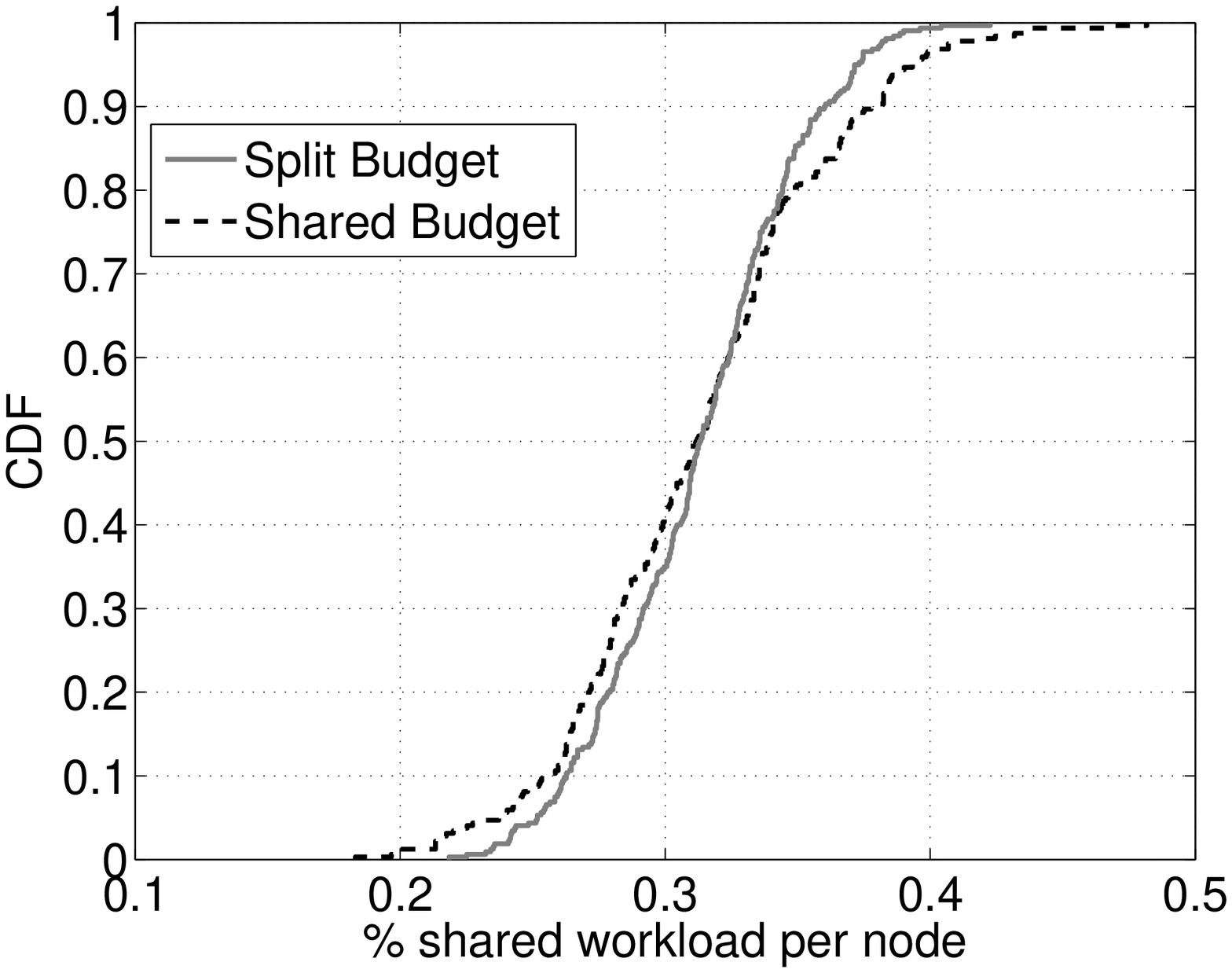}
}
\caption{CDFs of the stored items and of the workload at each network node}
\label{fig:performance-joint-vs-independent-items}
\vspace{-3mm}
\end{figure}

The scheme we propose is fair in terms of resources demanded from nodes
in the network. On the one hand, in Fig.~\ref{fig:cdf_item-joint-vs-independent},
we show the distribution of the number of items stored by a node at
the same time: a node seldom stores more than one replica, which implies
that node memory utilization is similar across the network.
Indeed, our scheme successfully avoids the risk of replica stacking at some
good candidates thanks to the enforced periodic swapping of the
replica role among nodes.
On the other hand, Fig.~\ref{fig:cdf_load-joint-vs-independent} depicts the
cumulative distribution function (CDF) of the percentage of total network workload
handled by each node, in terms of answered queries: the curve is quite
steep around the ideal value $\frac{1}{V}$ = 0.3\%, corresponding to a perfectly fair
workload distribution among nodes.

\subsection{Impact of the content characteristics}
\label{sec:multi-content-nat}

We now vary the popularity and size of content items, and observe their
impact on the performance of our replication scheme.

\begin{table}[tbp]
\centering
\caption{$R^*_i$ computed by the centralized CFL algorithm in presence
  of different content popularity\label{tab:loadbp-pop}}
      {
      \begin{tabular}{|c|c|c|c|}
        \hline
        {Item id} & {Interested} & {Opt. with split budget} &
        {Opt. with shared budget} \\
        \hline
        \hline
        {1} & {100\%} & {39} & {42} \\
        \hline
        {2} & {75\%} & {30} & {29} \\
         \hline
        {3} & {50\%} & {19} & {18} \\
        \hline
        {4} & {25\%} & {14} & {15} \\
        \hline
      \end{tabular}      
    } 
\end{table}

\begin{table}[t]
  \centering
\caption{$R^*_i$ computed by the centralized CFL algorithm in presence of
different content sizes\label{tab:loadbp-fs}}
      {
      \begin{tabular}{|c|c|c|c|}
        \hline
        {Item id} & {Item size} &  {Opt. with split budget} &
        {Opt. with shared budget} \\
        \hline
        \hline
        {1} & {1~Mbytes} & {39} & {42} \\
        \hline
        {2} & {2~Mbytes} & {62} & {67} \\
         \hline
        {3} & {3~Mbytes} & {87} & {91} \\
        \hline
        {4} & {4~Mbytes} & {115} & {117} \\
        \hline
      \end{tabular}      
    } 
\end{table}

\begin{figure*}[t]
  \begin{tabular}{ccc}

    \begin{minipage}[t]{0.3\textwidth}
      \begin{center}    
        \subfigure[Number of replicas (shared budget)]{
          \label{fig:rep-joint-multi-pop}
          \includegraphics[scale=0.25]{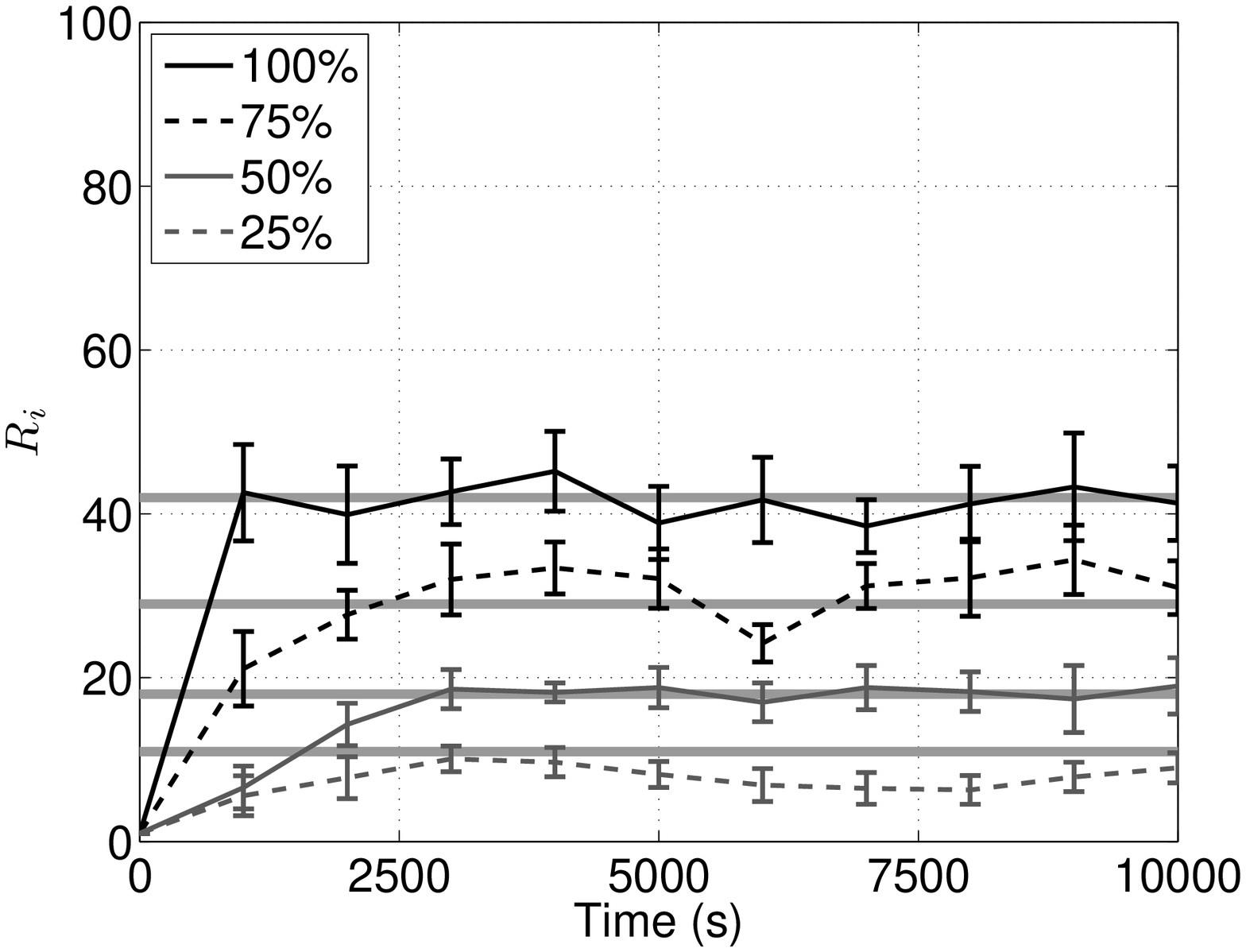}}
      \end{center}
    \end{minipage}

    &

    \begin{minipage}[t]{0.3\textwidth}
      \begin{center}    
        \subfigure[Total workload (shared budget)]{
          \label{fig:loadpernode_pop}
          \includegraphics[scale=0.25]{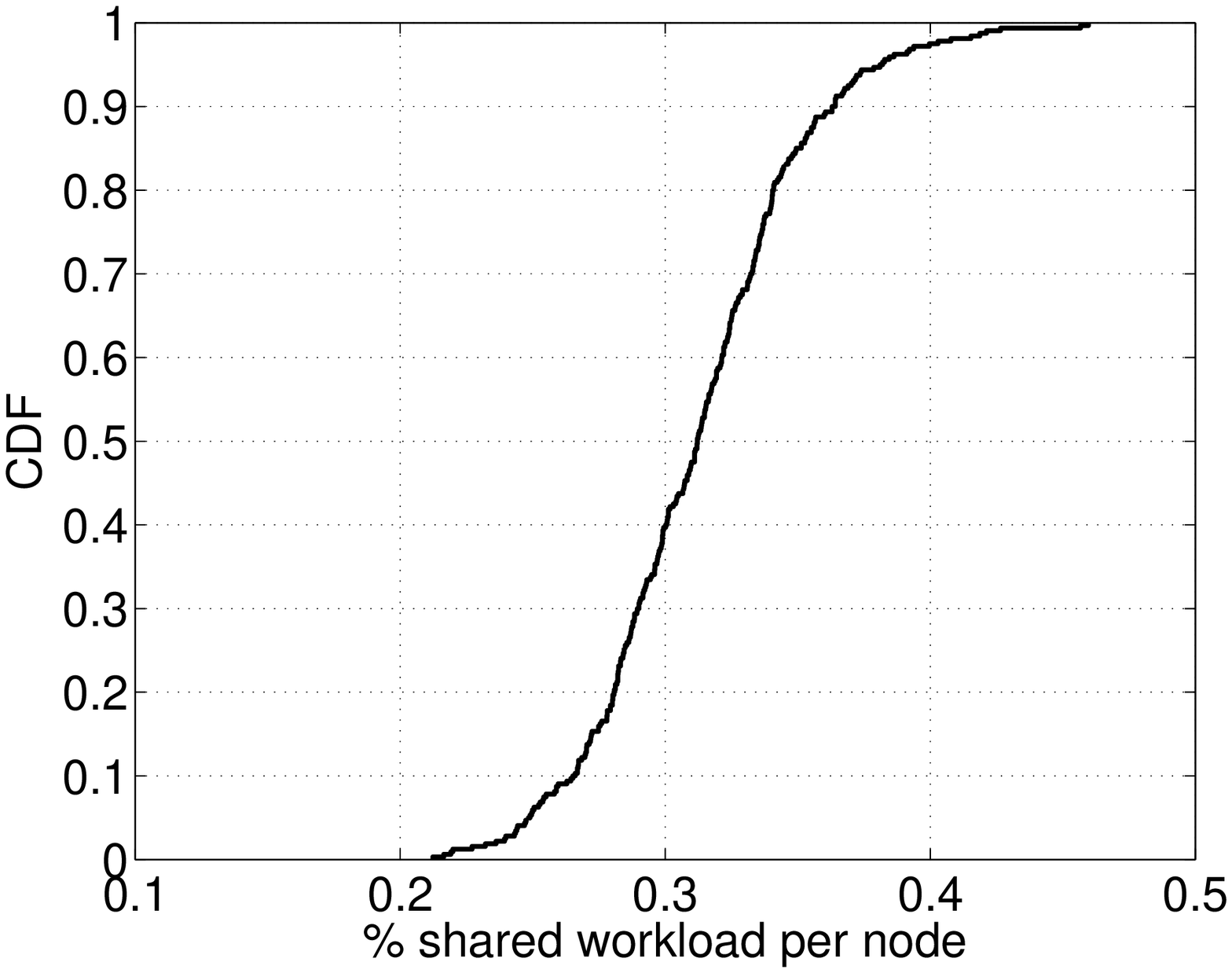}}
      \end{center}
    \end{minipage}

   &

    \begin{minipage}[t]{0.3\textwidth}
      \begin{center}    
        \subfigure[Stored items (shared budget)]{
          \label{fig:item-joint-multi-pop}
          \includegraphics[scale=0.25]{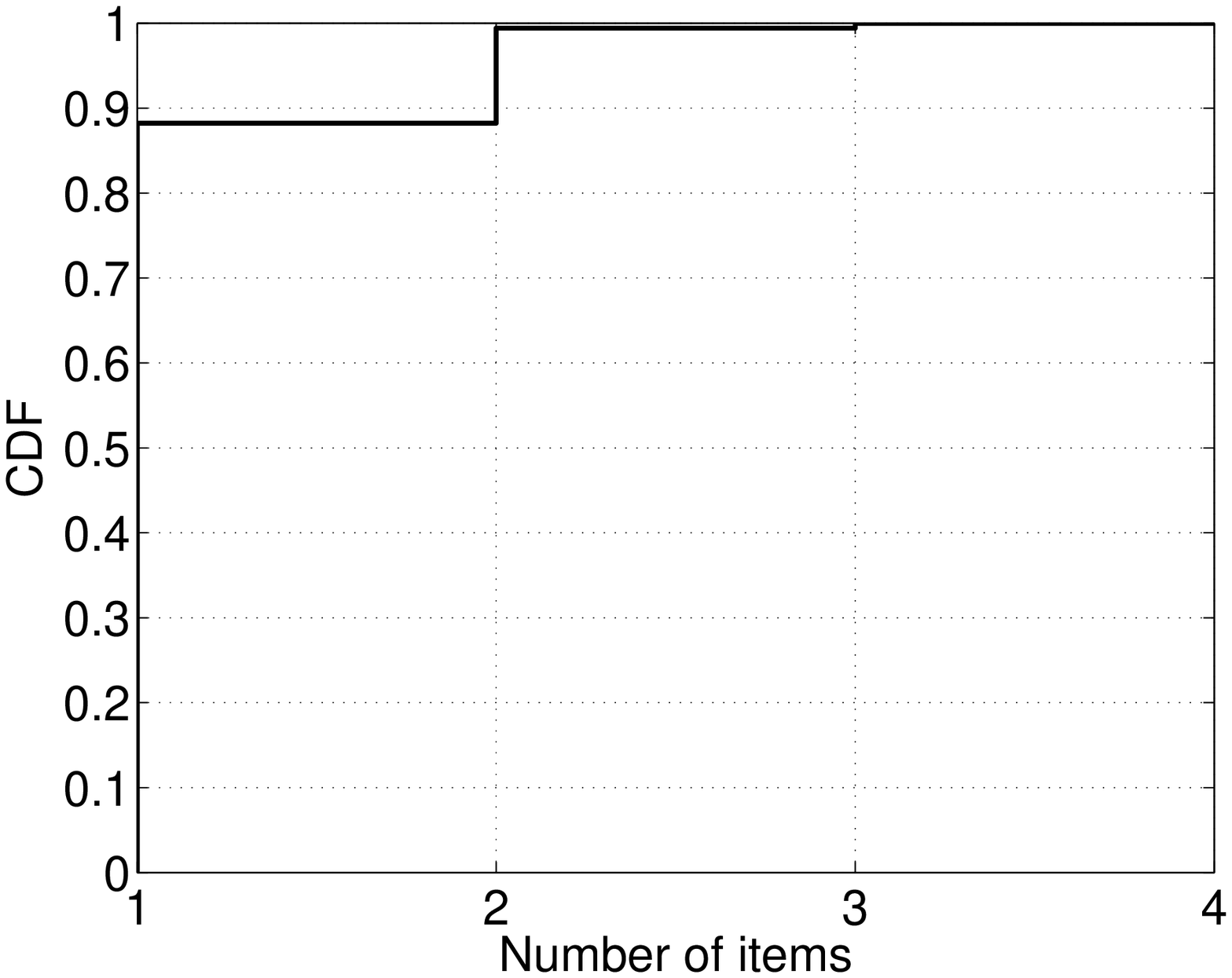}}
      \end{center}
    \end{minipage}




    
  \end{tabular}
  \caption{Impact of content popularity on the replication with
		shared capacity, in terms of number replicas, workload
    distribution, and memory utilization}
  \label{fig:performance-joint-multi-pop}
\vspace{-3mm}
\end{figure*}

\subsubsection*{How does our replication scheme perform 
in presence of items with different popularity?}
We study now the scenario when not all nodes are interested in a
content. In such a situation, a node stores a replica of the content only  
if it is interested. If a node attempts to hand over the content to an
uninterested node (by random selection), the request will be denied and
a different node will have to be selected. 

In Table~\ref{tab:loadbp-pop}, we report the results of the CFL
algorithm when the percentage of interested nodes, $p(i)$, $i=1,\ldots
4$, varies from 25\% to 100\%. We also set $c_{j}=40$~Mbytes for the
optimization with shared capacity budget and $c_j=60$~Mbytes for the
optimization with split capacity budget.
Interestingly, Table~\ref{tab:loadbp-pop} indicates that, in order for
the replication mechanisms to yield roughly the same replication
factor, the capacity budget that is required for the shared capacity
approach is substantially lower than that required for the split
capacity case.

As far as the optimization with shared capacity budget is concerned,
Fig.~\ref{fig:rep-joint-multi-pop} shows that the average number of
replicas for item $i$, $\bar{R}_i$, generated by our scheme oscillates
around the optimal value determined by the CFL algorithm for the same
item, $R^*_i$, even when $i$ is characterized by low popularity.
Moreover, the workload remains evenly shared
among replica nodes: Fig.~\ref{fig:loadpernode_pop} shows that each
node serves at least 0.2\% of the total workload and 98\% of nodes
serve less than 0.4\% of the total workload. The load distribution is
thus quite dense around 0.3\%, i.e., $\frac{1}{V}$ that is the ideal mean
workload. Finally, the results in Fig.~\ref{fig:item-joint-multi-pop}
underline the fairness of our
replication scheme also from a memory utilization  point of view, with
nodes caching with high probability at most one content at a time. 
We observe similar results\footnote{For the sake of brevity,
we omit these results in this work.}
when the split capacity approach is used, although this requires a larger 
budget to be allocated to the replication process. 


\begin{figure*}[t]
  \begin{tabular}{ccc}

    \begin{minipage}[t]{0.3\textwidth}
      \begin{center}    
        \subfigure[Number of replicas (shared budget)]{
          \label{fig:rep-joint-multi-fs}
          \includegraphics[scale=0.25]{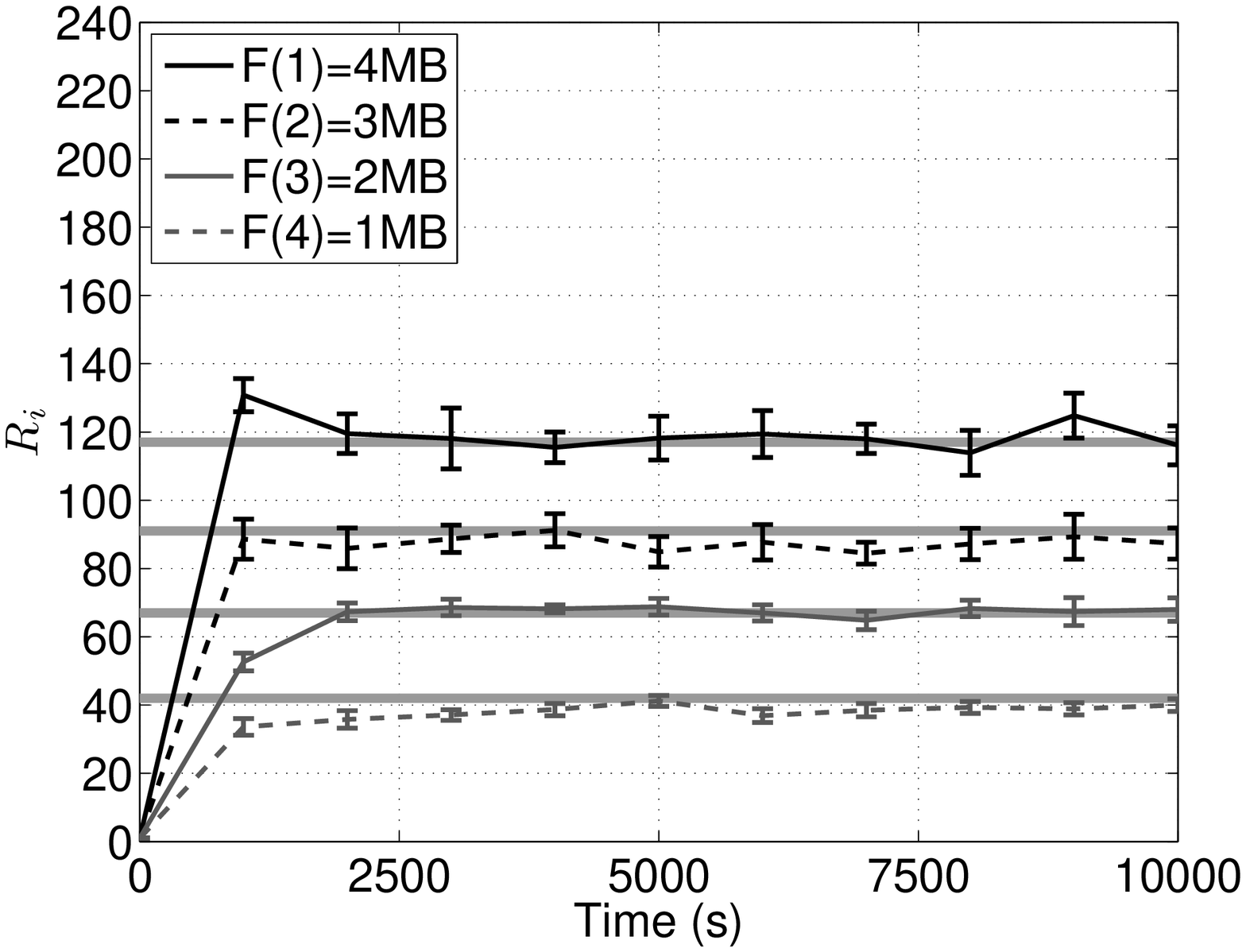}}
      \end{center}
    \end{minipage}

    &

    \begin{minipage}[t]{0.3\textwidth}
      \begin{center}    
        \subfigure[Total workload (shared budget)]{
          \label{fig:loadpernode_fs}
          \includegraphics[scale=0.25]{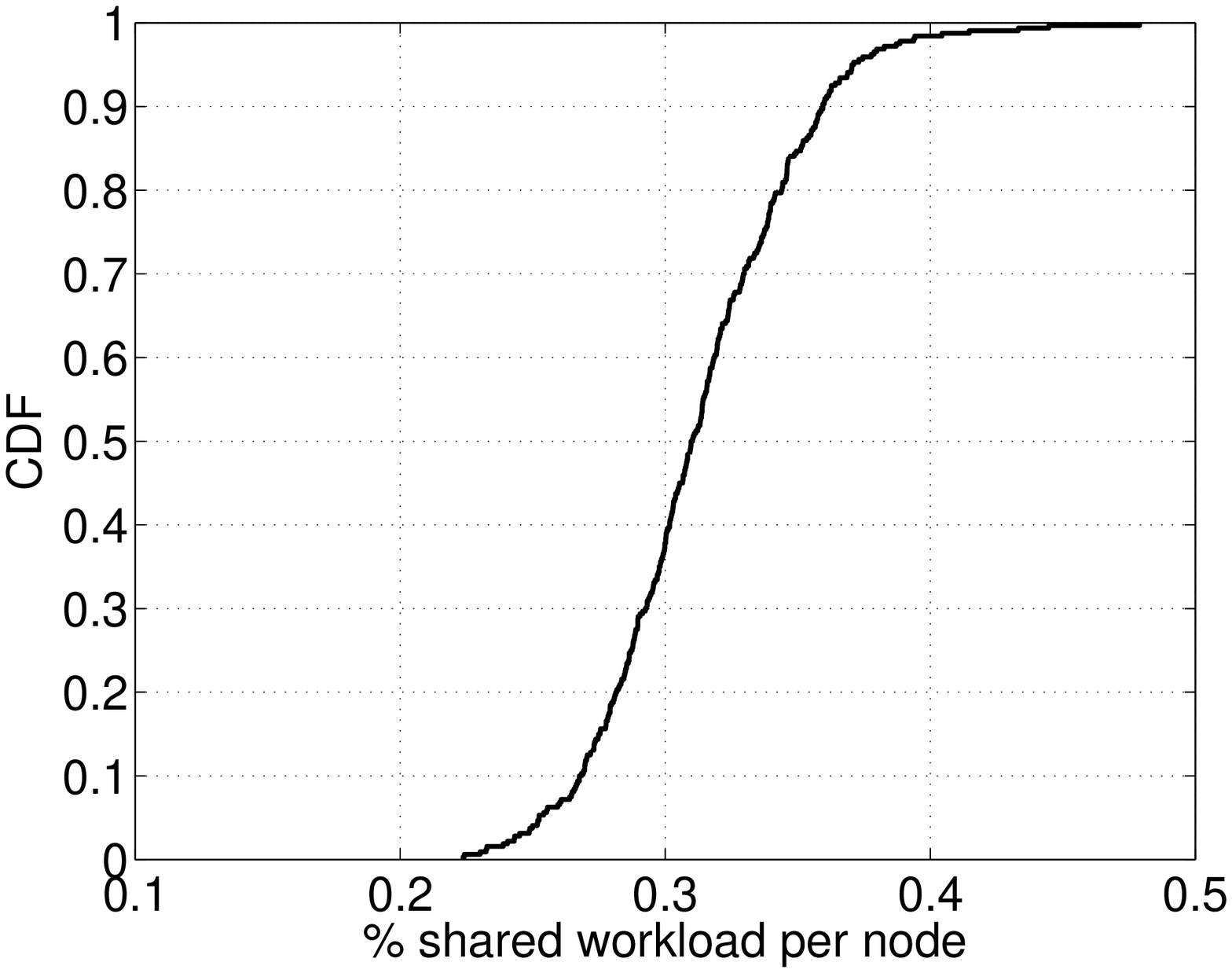}}
      \end{center}
    \end{minipage}

   &

    \begin{minipage}[t]{0.3\textwidth}
      \begin{center}    
        \subfigure[Stored items (shared budget)]{
          \label{fig:item-joint-multi-fs}
          \includegraphics[scale=0.25]{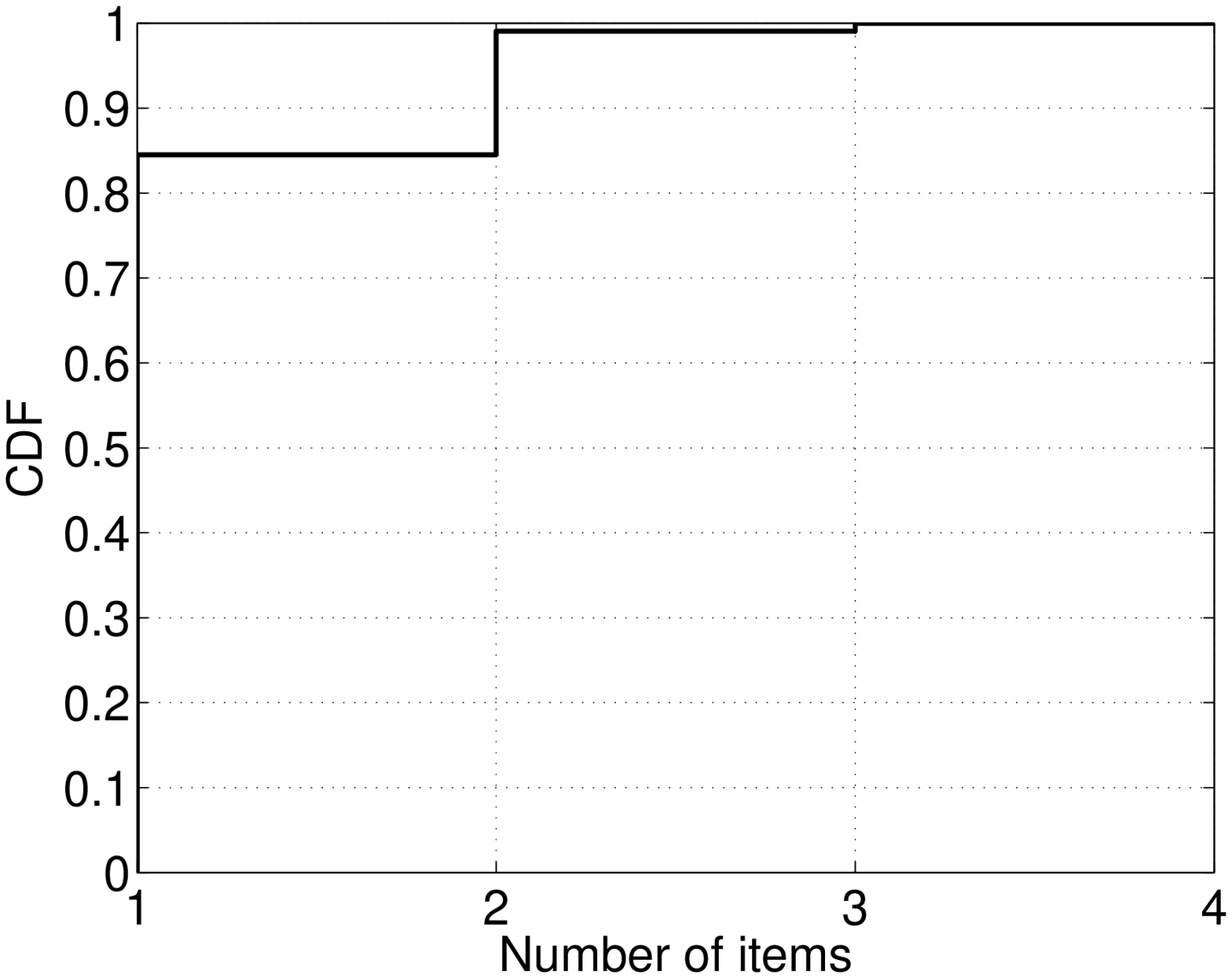}}
      \end{center}
    \end{minipage}




     
  \end{tabular}
  \caption{Impact of content size on the replication with
		shared capacity, in terms of number of replicas, workload
    distribution, and memory utilization}
  \label{fig:performance-joint-multi-fs}
\vspace{-3mm}
\end{figure*}

\subsubsection*{How does our replication scheme perform with
different content sizes?}
Let us focus on a scenario where the four items have identical popularity
but different sizes ($s(i)$, $i=1,\ldots,4$). The considered values
are detailed in Table~\ref{tab:loadbp-fs}, along with the optimal
number of replicas $R^*_i$ computed, for each item, by the centralized
CFL algorithm under the split and shared capacity budget
constraints. 

Focusing on the optimization problem with shared capacity budget, 
Fig.~\ref{fig:rep-joint-multi-fs} shows a good matching between
$\bar{R}_i$ and the optimal value  $R^*_i$, for any item $i$. 
The workload exacted from the nodes by our scheme is shown in
Fig.~\ref{fig:loadpernode_fs}, and the number of information items
stored by each node is depicted in Fig.~\ref{fig:item-joint-multi-fs}.
Very similar considerations apply to the case of optimization with
split capacity budget, 
although comparable performance can only be attained if the capacity
budget
allocated by each node largely exceeds that in the shared capacity approach.

%

\begin{figure*}[t]
  \begin{tabular}{ccc}
    \begin{minipage}[t]{0.3\textwidth}
      \begin{center}    
        \subfigure[Number of replicas (shared budget)]{
          \label{fig:rep_joint_multi_slaw}
          \includegraphics[scale=0.25]{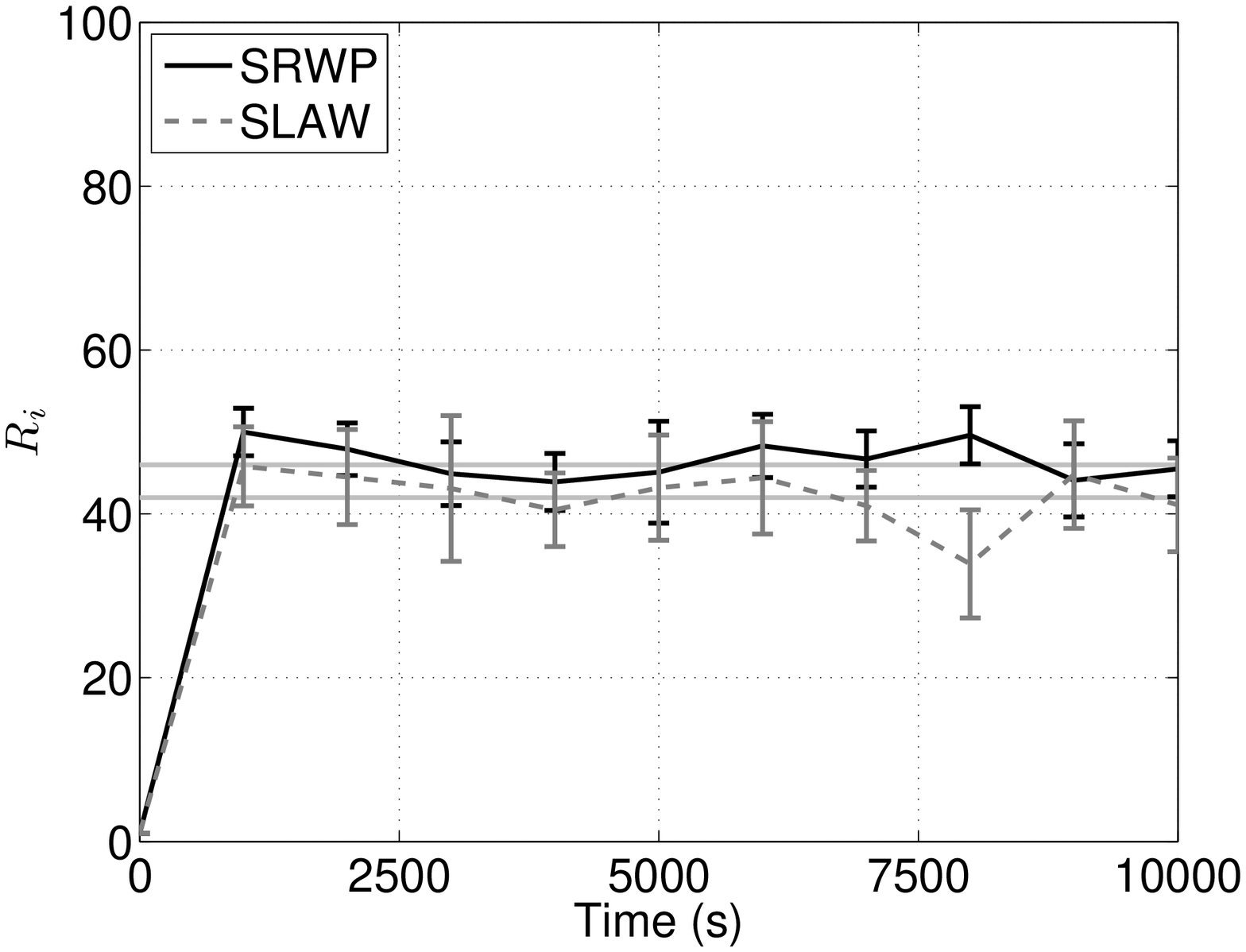}}
      \end{center}
    \end{minipage}

    &

    \begin{minipage}[t]{0.3\textwidth}
      \begin{center}    
        \subfigure[Total workload (shared budget)]{
          \label{fig:loadpernode_slaw}
          \includegraphics[scale=0.25]{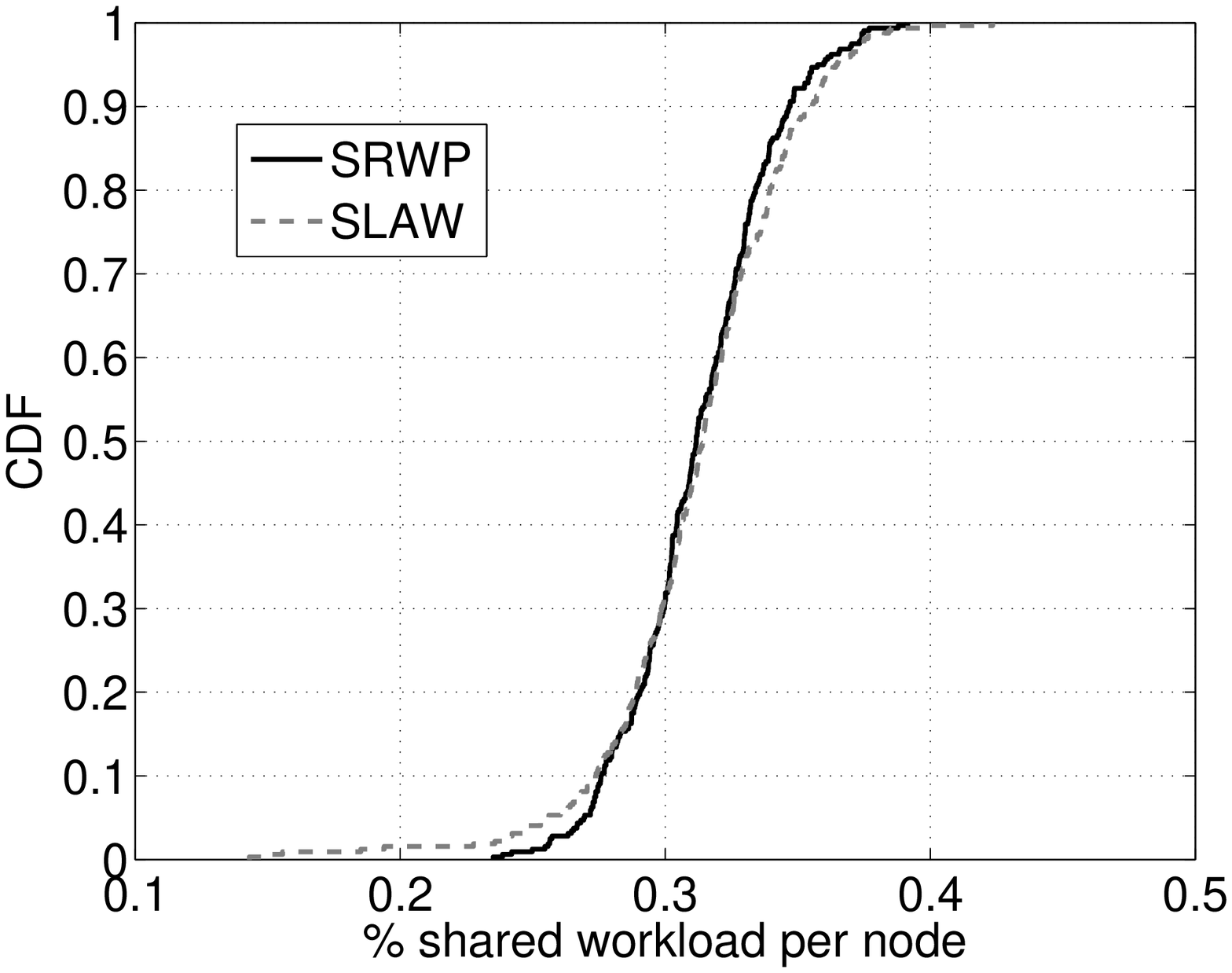}}
      \end{center}
    \end{minipage}
	
  &	
     \begin{minipage}[t]{0.3\textwidth}
      \begin{center}    
        \subfigure[Stored items (shared budget)]{
          \label{fig:item_joint_multi_slaw}
          \includegraphics[scale=0.25]{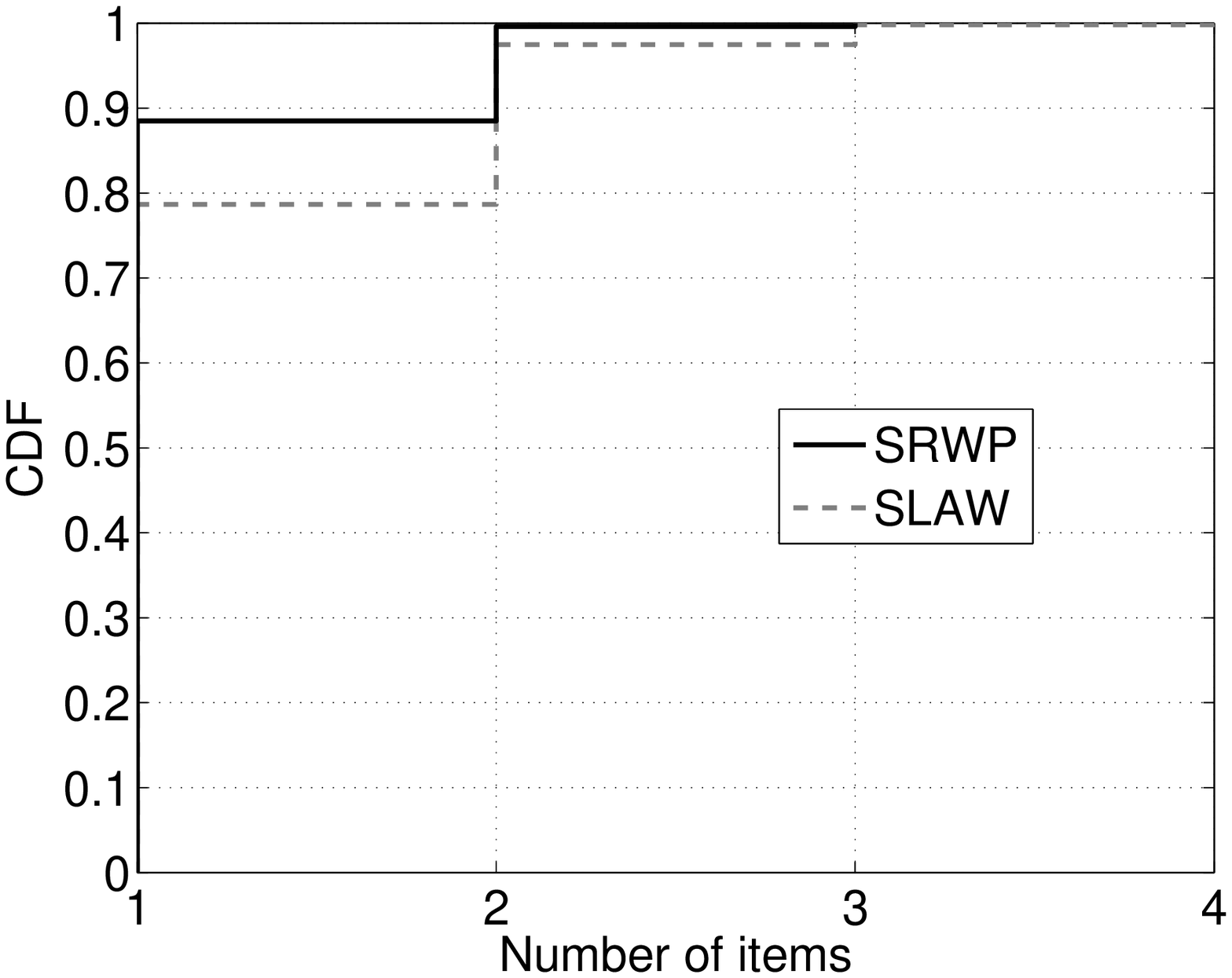}}
      \end{center}
    \end{minipage}

  \end{tabular}
  \caption{Impact of user mobility on the replication with
		shared capacity, in terms of number of replicas, workload
    distribution, and memory utilization}
  \label{fig:performance-joint-multi-slaw}
\vspace{-3mm}
\end{figure*}

\begin{figure*}[t]
  \begin{tabular}{ccc}

    \begin{minipage}[t]{0.3\textwidth}
      \begin{center}    
        \subfigure[Number of replicas (shared budget)]{
          \label{fig:replica_velocity}
          \includegraphics[scale=0.25]{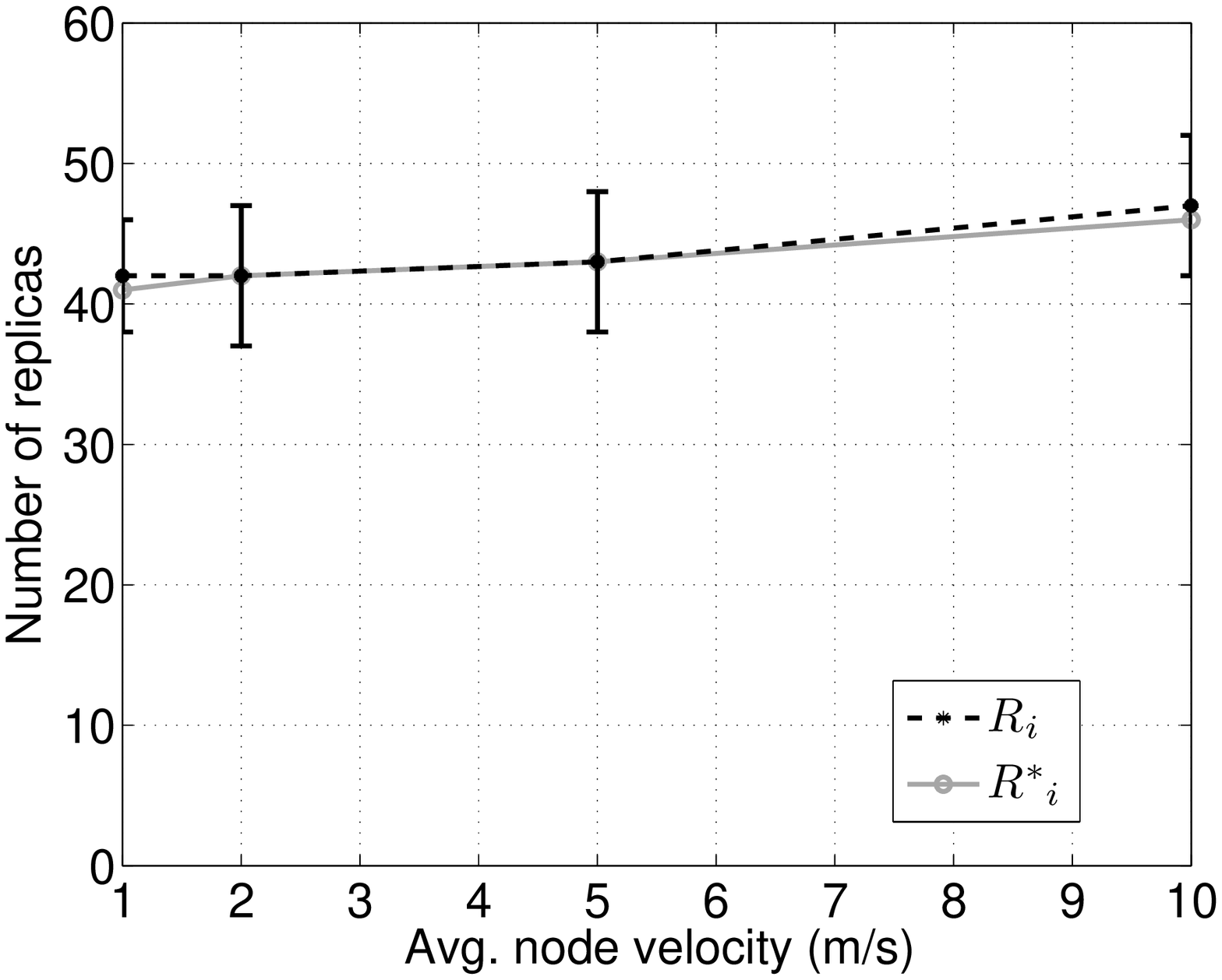}}
      \end{center}
    \end{minipage}

    &

    \begin{minipage}[t]{0.3\textwidth}
      \begin{center}    
        \subfigure[Workload (shared budget)]{
          \label{fig:load_velocity}
          \includegraphics[scale=0.25]{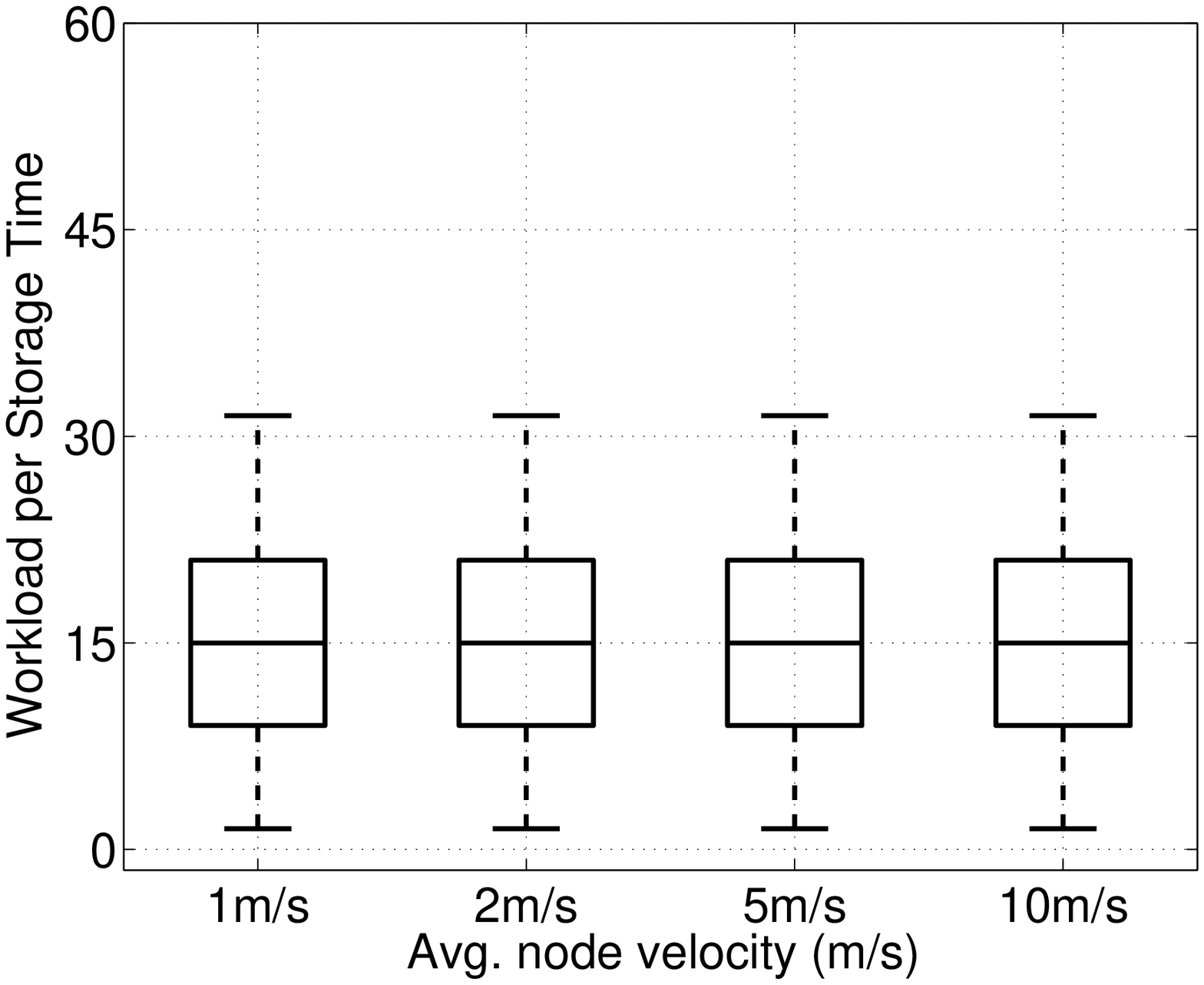}}
      \end{center}
    \end{minipage}
	
  &	
     \begin{minipage}[t]{0.3\textwidth}
      \begin{center}    
        \subfigure[Delay (shared budget)]{
          \label{fig:delay_velocity}
          \includegraphics[scale=0.25]{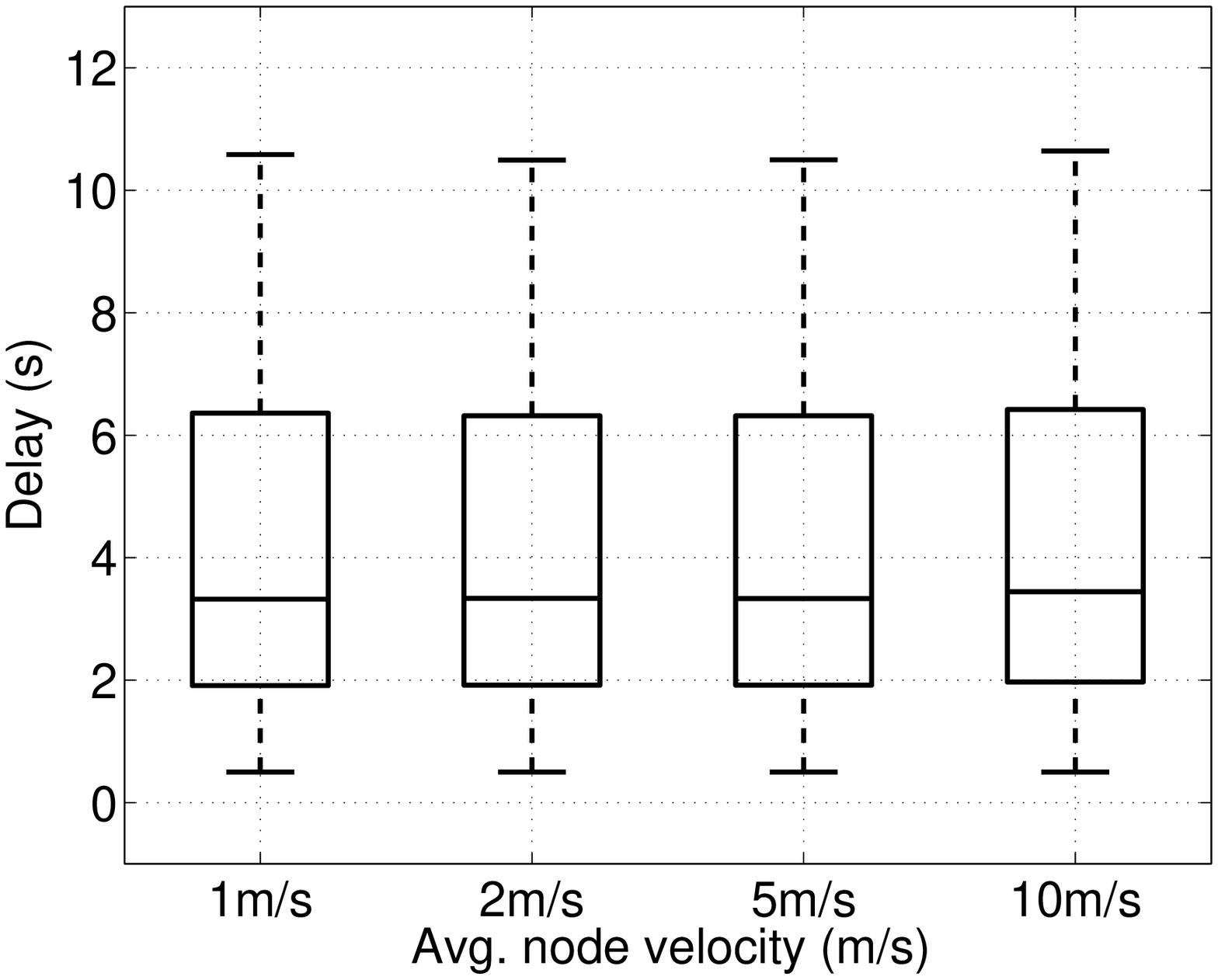}}
      \end{center}
    \end{minipage}

  \end{tabular}
  \caption{Impact of node speed on the replication with
		shared capacity, in terms of number of replicas, workload
    distribution, and memory utilization}
  \label{fig:performance_velocity}
\vspace{-3mm}
\end{figure*}

\subsection{Impact of mobility}

\subsubsection*{What is the impact of a more accurate human mobility
  model on our scheme?}
\label{sec:syntheticmobility}
We now study the performance of our scheme in presence of non-random
clustered mobility, which has been shown to characterize human
movements in outdoor environments. 
More precisely, we employed the SLAW model~\cite{slaw} to generate
a synthetic trace representing the movements of 320 outdoor users
within an area of 1~km$^2$, during 3 hours.
The SLAW settings included 600 waypoints, Pareto-distributed with Hurst
parameter equal to 0.75, a flight speed of~1 m/s, and pause times that
obey a Levy distribution with coefficient equal to 1 and minimum
and maximum values equal to 100~s and 1000~s, respectively.
The distance weight, which determines the priority that nodes give to
nearby locations before traveling to farther locations, is set to 3.
All results refer to the case of the optimization with shared capacity
budget: those for the optimization problem with split capacity budget
are very similar and are omitted for sake of brevity.

Fig.~\ref{fig:rep_joint_multi_slaw} shows the evolution of the number
of replicas per information item over the simulation time, for SLAW
and the stationary RWP previously employed. In both cases, the number
of replicas per item roughly matches the optimal value. 
In the SLAW scenario, the presence of a small number of dense clusters
implies that content queries will be originating from within each
cluster: this explains the (almost negligible) difference in the
number of replicas and workload with respect to the RWP model.
It also follows that the different mobility does
not result in significant differences in the total load
distribution, as shown by the plot in Fig.~\ref{fig:loadpernode_slaw}.
As far as memory utilization is concerned, in
Fig.~\ref{fig:item_joint_multi_slaw} 
SLAW forces a slightly more unbalanced CDF, as nodes group into denser 
clusters than under RWP mobility. Specifically,  under SLAW,
80\% of nodes hold two or more items versus the 90\% measured under the RWP model. 

\subsubsection*{How does our mechanism work as the node speed varies?}
\label{sec:speedmobility}

Invariance of the performance of our replication scheme to the
node speed is demonstrated by Fig.~\ref{fig:replica_velocity},
Fig.~\ref{fig:load_velocity} and Fig.~\ref{fig:delay_velocity}.
There, we can notice how the different velocity of nodes during their
movement does not lead to significant variations in the number of
replicas, per-node workload and delay, respectively.

\begin{figure*}[t]

  \begin{tabular}{ccc}
    \begin{minipage}[t]{0.3\textwidth}
      \begin{center}    
        \subfigure[Number of replicas (shared budget)]{
          \label{fig:replica_item}
          \includegraphics[scale=0.25]{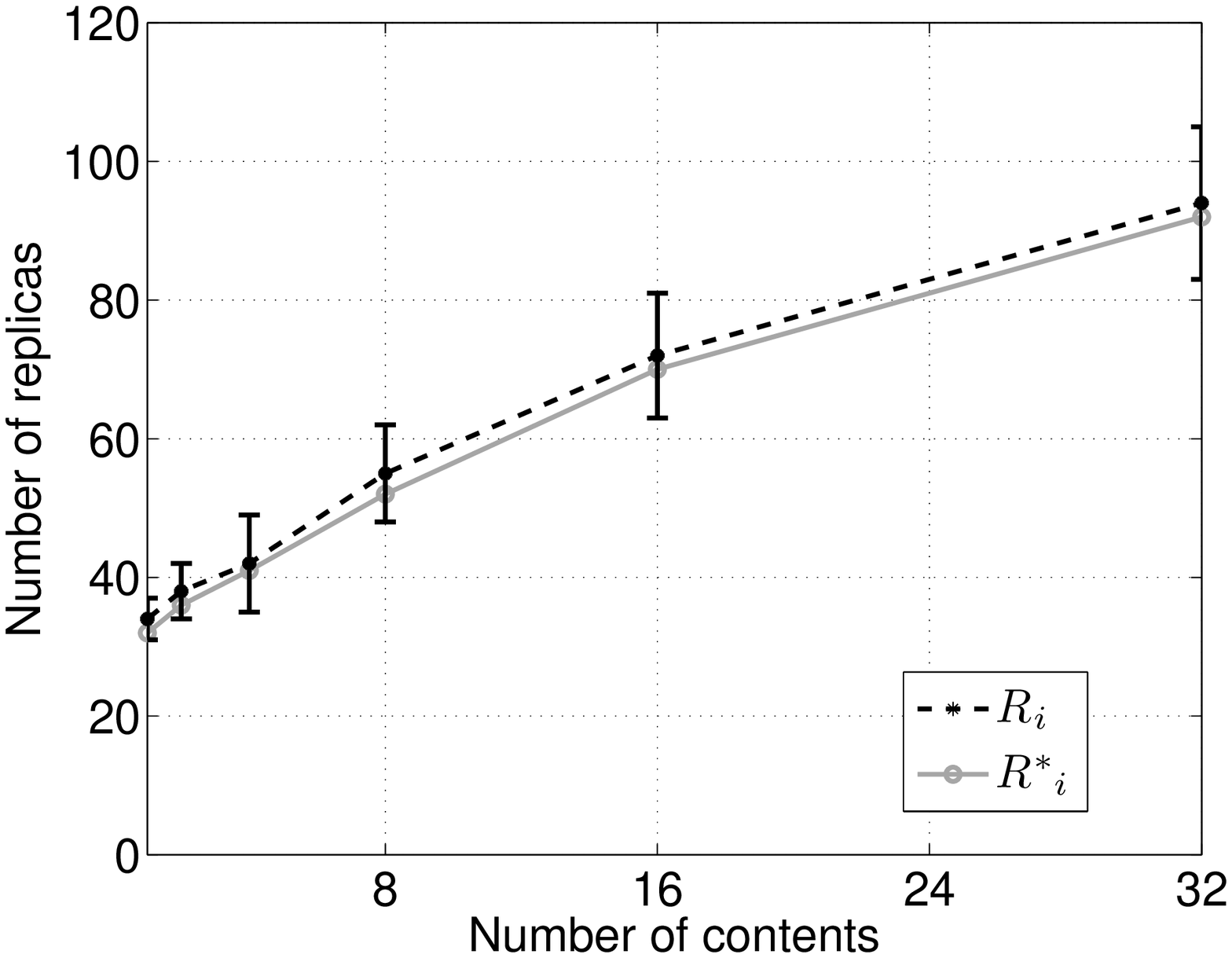}}
      \end{center}
    \end{minipage}

    &

    \begin{minipage}[t]{0.3\textwidth}
      \begin{center}    
        \subfigure[Workload (shared budget)]{
          \label{fig:load_item}
          \includegraphics[scale=0.25]{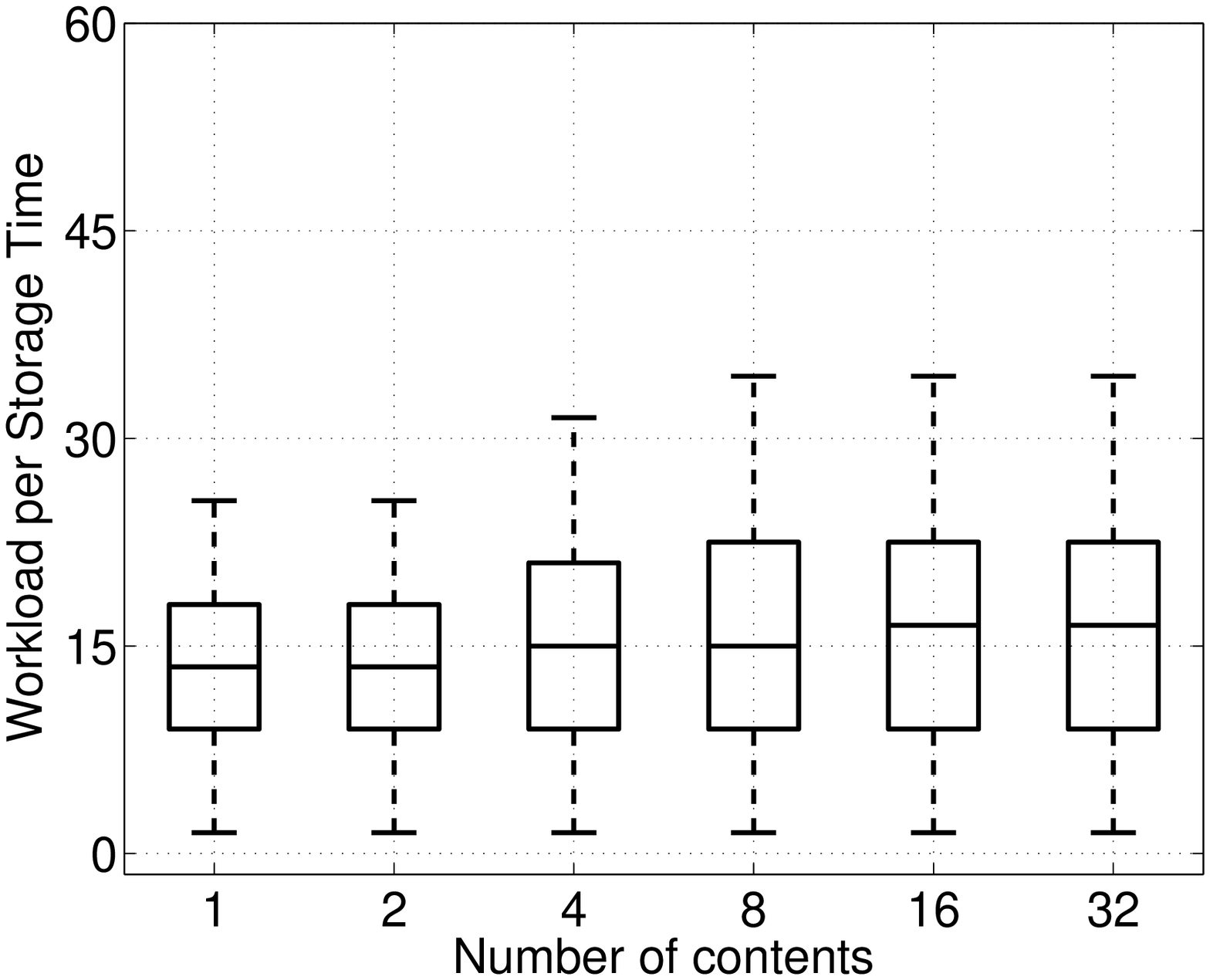}}
      \end{center}
    \end{minipage}
	
  &	
     \begin{minipage}[t]{0.3\textwidth}
      \begin{center}    
        \subfigure[Delay (shared budget)]{
          \label{fig:delay_item}
          \includegraphics[scale=0.25]{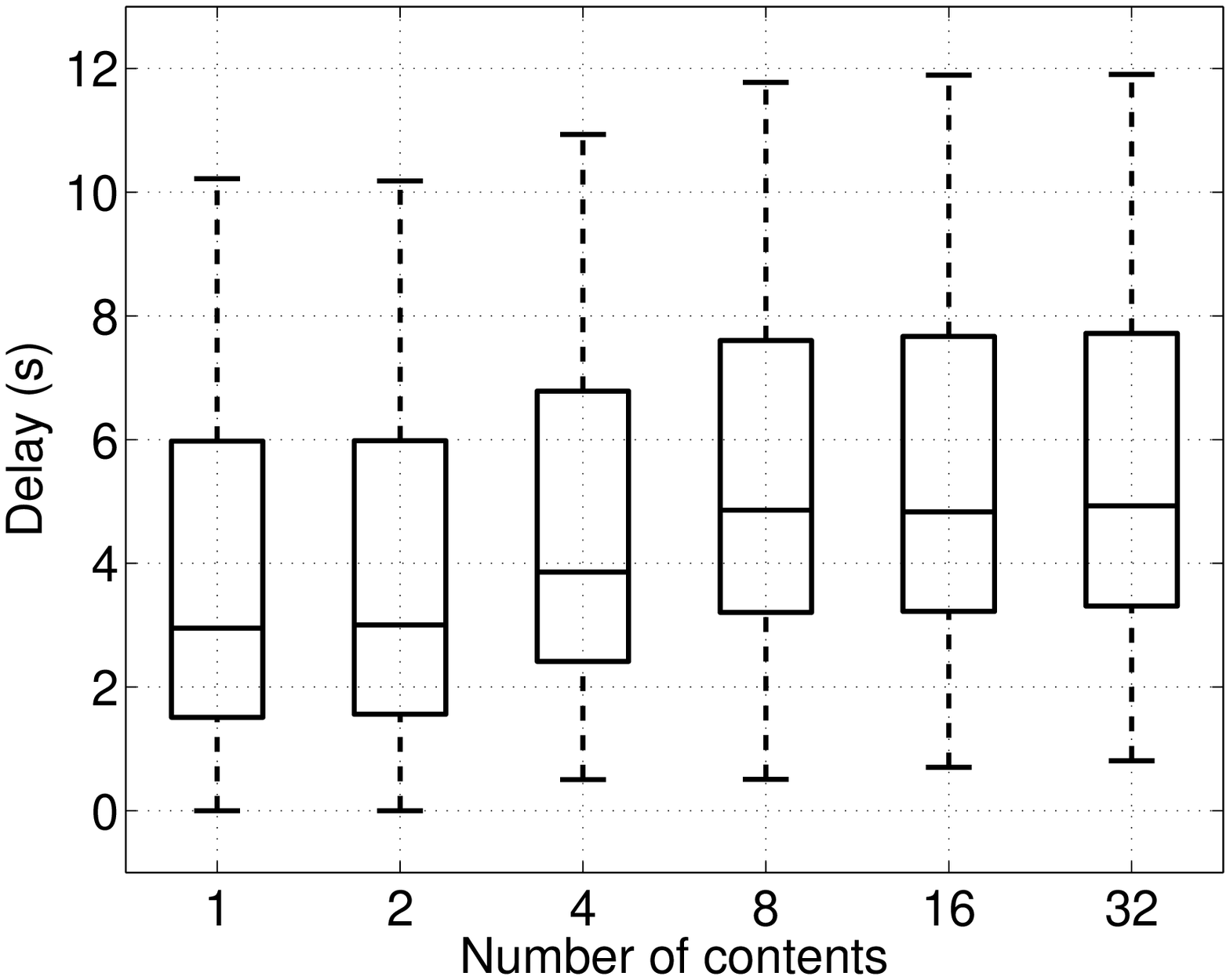}}
      \end{center}
    \end{minipage}

  \end{tabular}
  \caption{Impact of content set cardinality on the replication with
		shared capacity, in terms of number of replicas, workload
    distribution, and delay}
  \label{fig:performance_item}
\vspace{-3mm}
\end{figure*}

\begin{figure*}[t]
  \begin{tabular}{ccc}

    \begin{minipage}[t]{0.3\textwidth}
      \begin{center}    
        \subfigure[Number of replicas (shared budget)]{
          \label{fig:replica_deg}
          \includegraphics[scale=0.25]{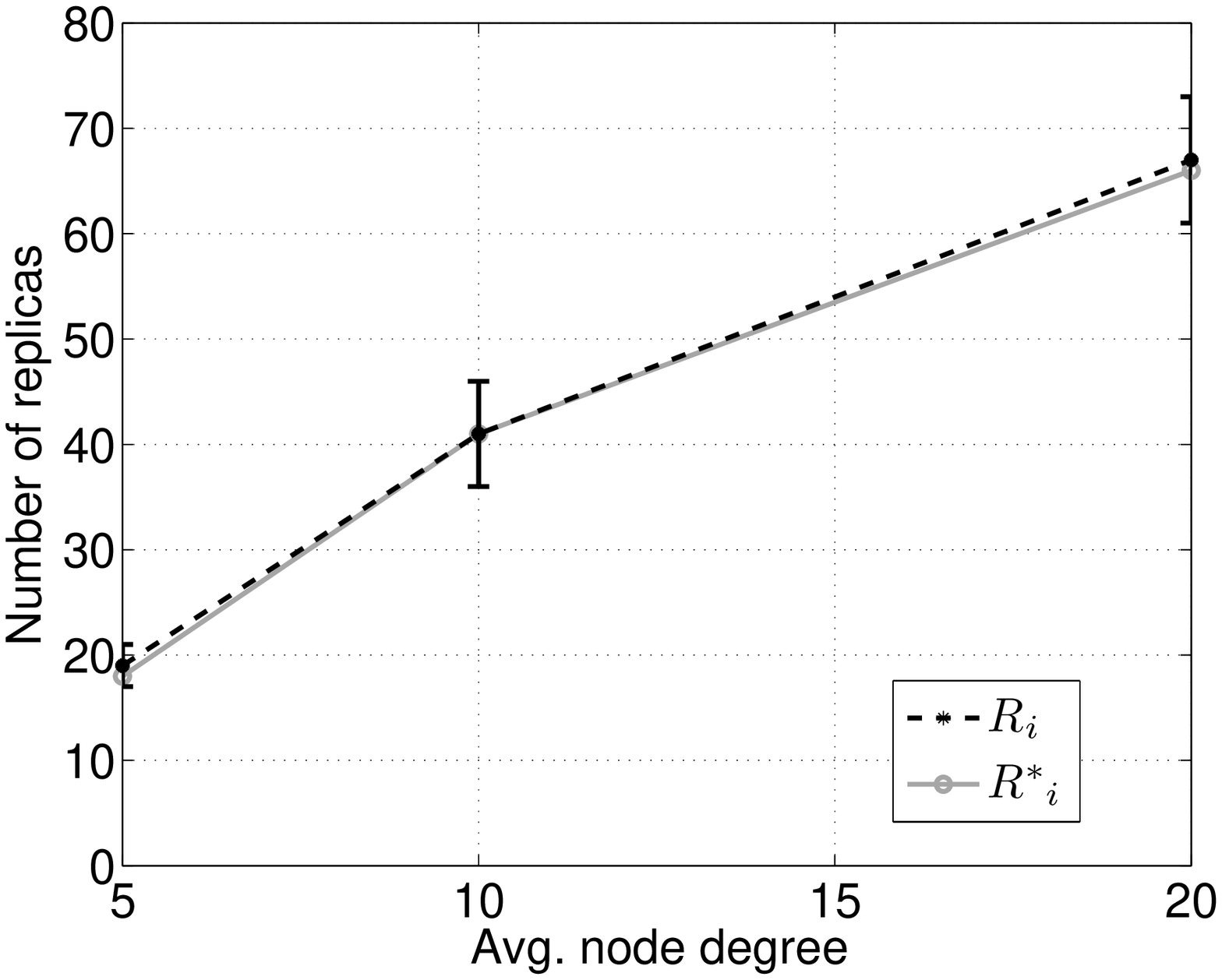}}
      \end{center}
    \end{minipage}

    &

    \begin{minipage}[t]{0.3\textwidth}
      \begin{center}    
        \subfigure[Workload (shared budget)]{
          \label{fig:load_deg}
          \includegraphics[scale=0.25]{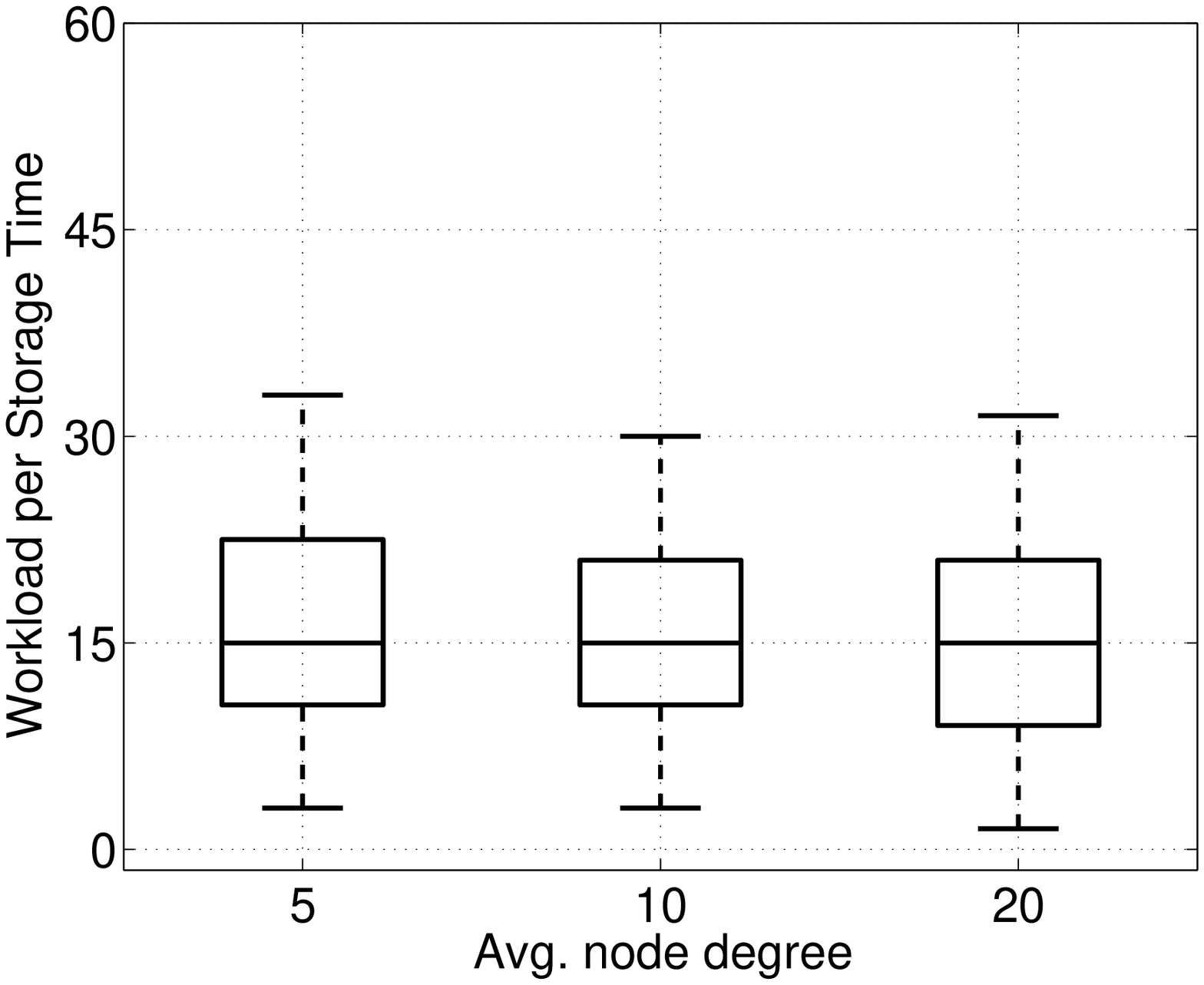}}
      \end{center}
    \end{minipage}
	
  &	
     \begin{minipage}[t]{0.3\textwidth}
      \begin{center}    
        \subfigure[Delay (shared budget)]{
          \label{fig:delay_deg}
          \includegraphics[scale=0.25]{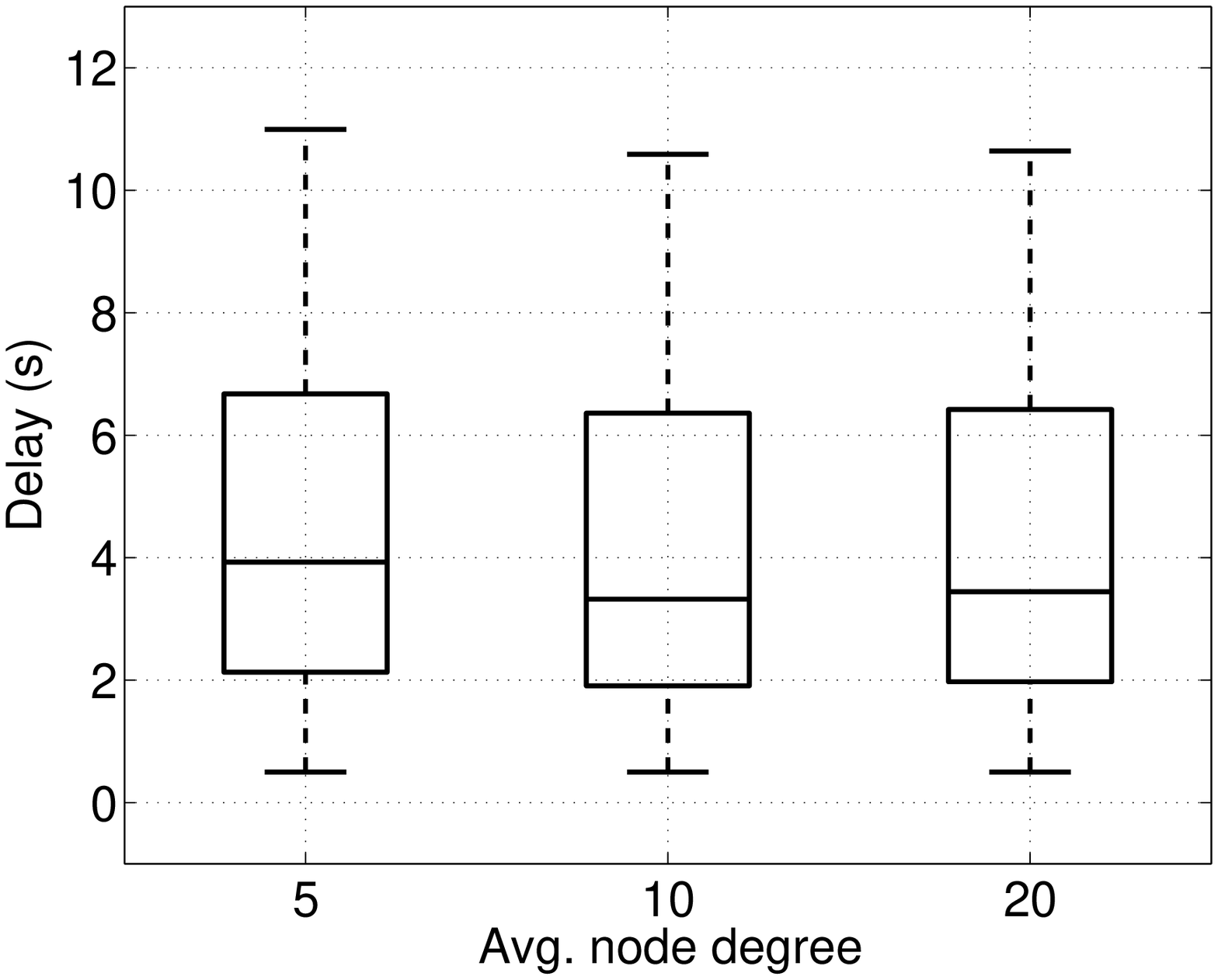}}
      \end{center}
    \end{minipage}

  \end{tabular}
  \caption{Impact of network density on the replication with
		shared capacity, in terms of number of replicas, workload
    distribution, and delay}
  \label{fig:performance_deg}
\vspace{-3mm}
\end{figure*}

\begin{figure*}[t]
  \begin{tabular}{ccc}
 
    \begin{minipage}[t]{0.3\textwidth}
      \begin{center}    
        \subfigure[Number of replicas (shared budget)]{
          \label{fig:rep_network_size}
          \includegraphics[scale=0.25]{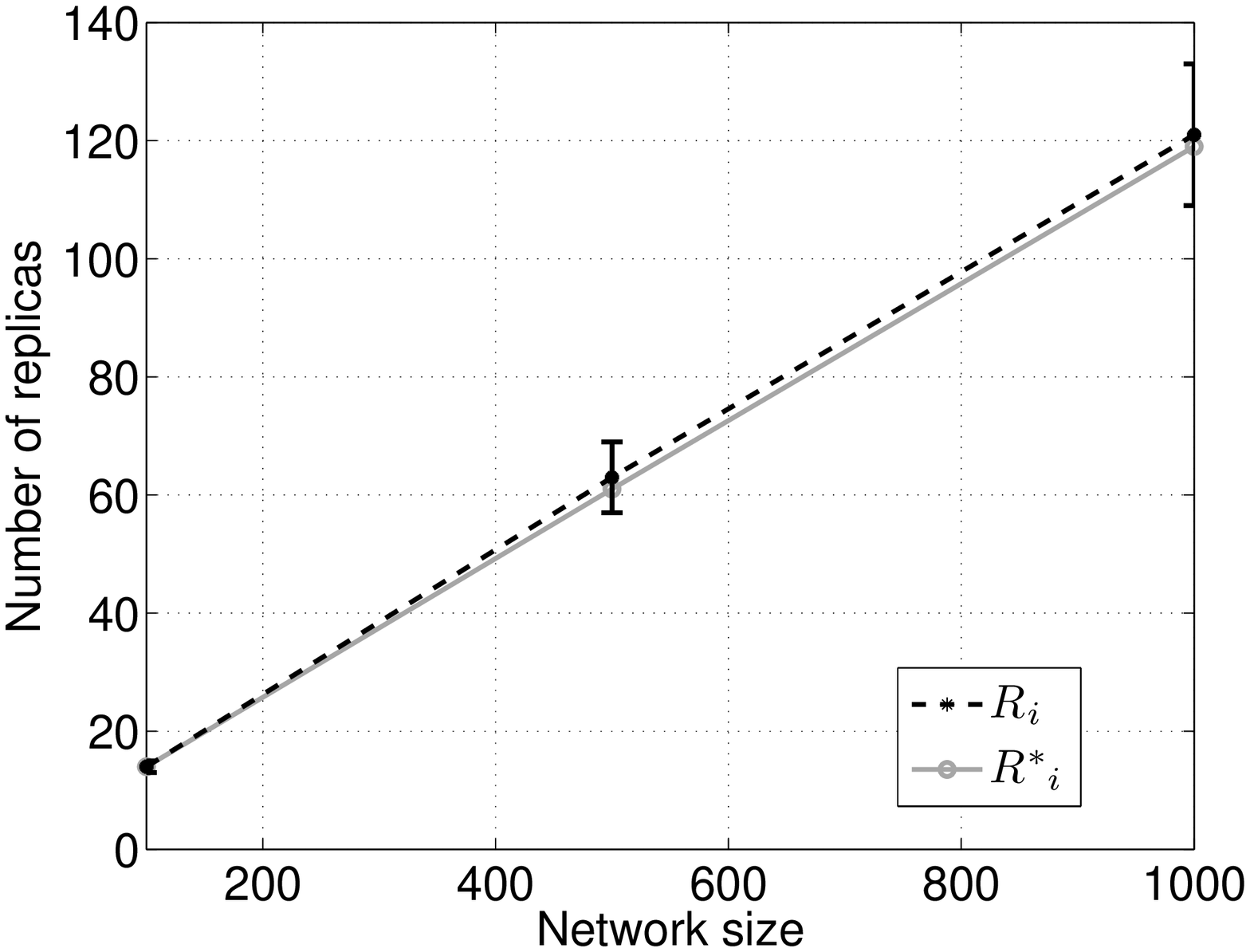}}
      \end{center}
    \end{minipage}

    &
    \begin{minipage}[t]{0.3\textwidth}
      \begin{center}    
        \subfigure[Workload (shared budget)]{
          \label{fig:load_network_size}
          \includegraphics[scale=0.25]{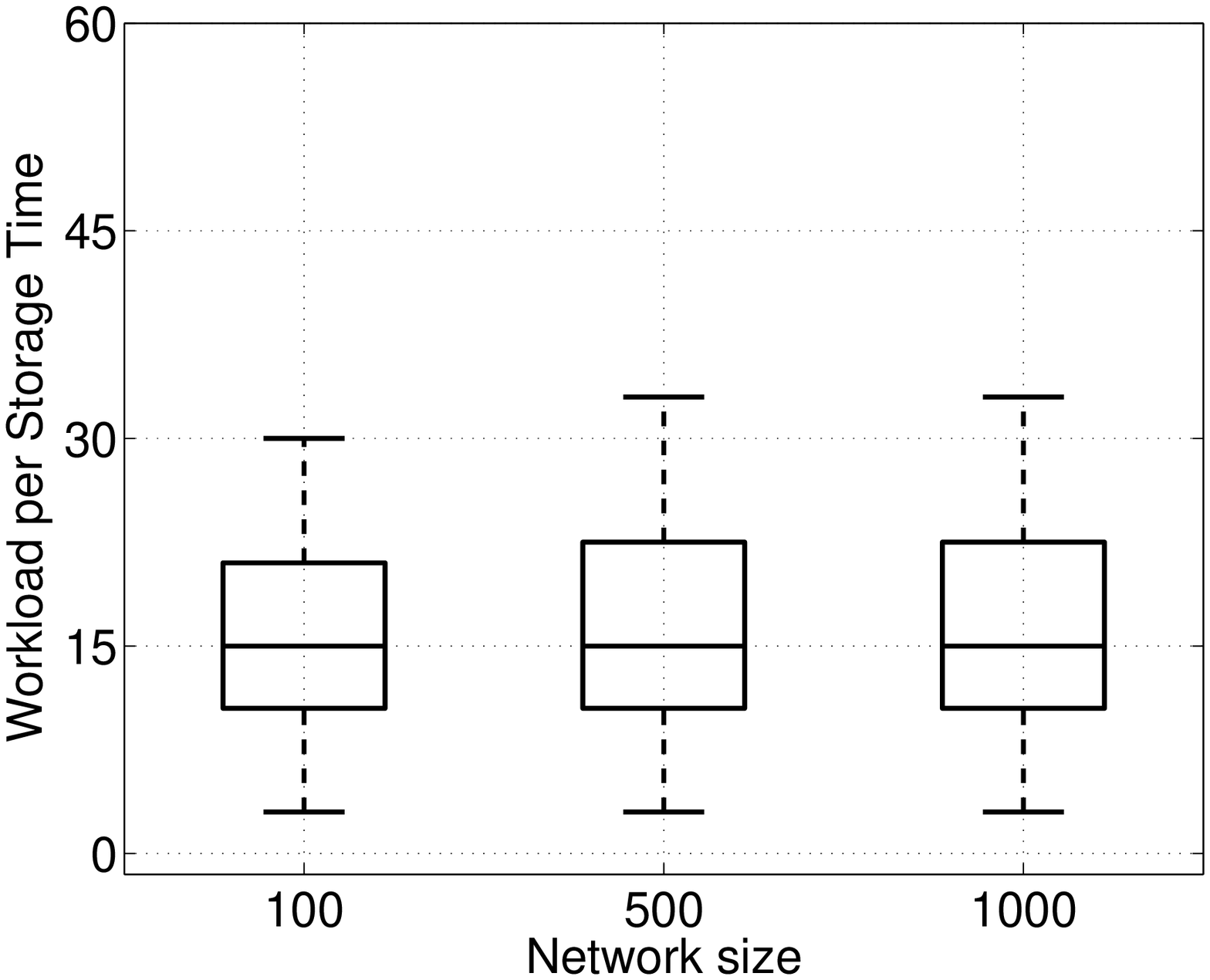}}
      \end{center}
    \end{minipage}

    &

 \begin{minipage}[t]{0.3\textwidth}
      \begin{center}    
        \subfigure[Delay (shared budget)]{
          \label{fig:delay_network_size}
          \includegraphics[scale=0.25]{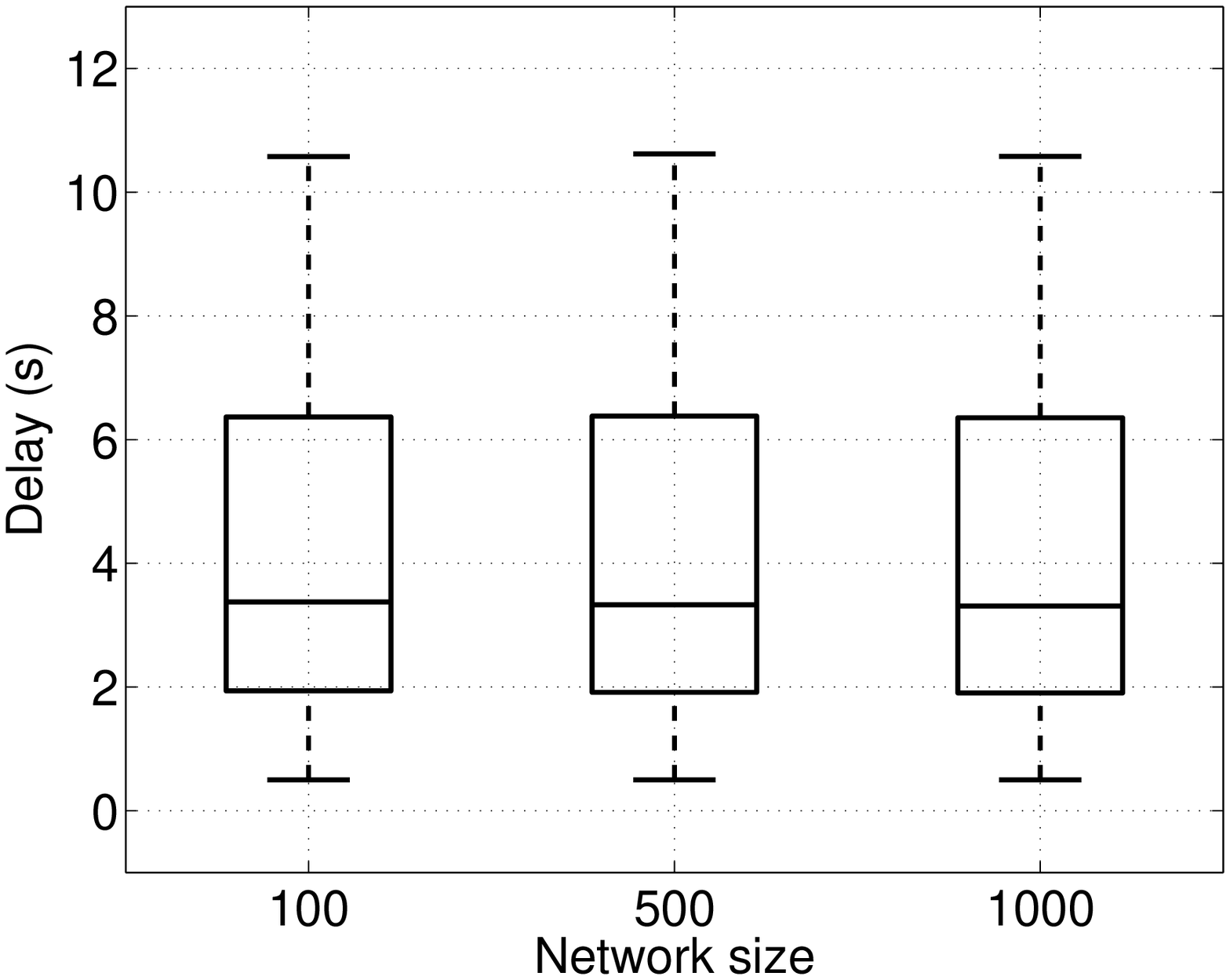}}
      \end{center}
    \end{minipage}

  \end{tabular}
  \caption{Impact of network size on the replication with
		shared capacity, in terms of number of replicas, workload
    distribution, and delay}
  \label{fig:per_network_size}
\vspace{-3mm}
\end{figure*}

\subsection{Scalability}
In order to determine the scalability properties of the proposed
replication scheme, we study the impact that the number of items,
network density, and network size have on the system
performance. Again, all results refer to the case of optimization with
shared capacity budget, since those obtained under optimization with
split capacity budget are similar, but require a significantly higher
budget to be allocated at nodes.

We first evaluate the performance when the cardinality of the
information item set varies between 1 and 32.  More precisely,
Fig.~\ref{fig:replica_item} shows the number of replicas per item
generated in the system, which grows as the size of the information 
set increases. Indeed, a larger content set implies that nodes tend to store
more items on average; however, their capacity budget $c_j$
remains constant, and is shared among all items they store. As a
result, focusing on one single content, each replica node for that
content will be able to serve fewer and fewer queries as the number of
available items increases. As a consequence, 
more replicas for the same content are
needed in order to meet the constraint on the capacity budget, hence to  
keep the workload constant, as depicted in
Fig.~\ref{fig:load_item}. 

Fig.~\ref{fig:delay_item} shows the effect that the number of
information item has on the service provisioning delay. The increase
of the delays is imputable to the heavier traffic on the channel, that
results in collisions and retransmissions of the information replies. 

We then study the effect of the network density, measured
as the average node degree, which is increased up to a mean number of 
neighbors per node equal to  20 in Fig.~\ref{fig:performance_deg}.  
Fig.~\ref{fig:replica_deg} shows that the number of replicas increases
according to the optimal number of facilities computed by the CFL
local search algorithm. 
Indeed, the increased presence of neighbors
induces a higher load in the network, in terms of queries: in order
to satisfy the new demand, and yet fulfill the per-node workload
constraint, additional nodes must become providers for each content.
The availability of additional facility nodes allows them to experience
a practically unchanged per-storage time workload, in
Fig.~\ref{fig:load_deg}, as well as a similar delay for successful
content requests, in Fig.~\ref{fig:delay_deg}.

Finally, in Fig.~\ref{fig:per_network_size}, we assess the performance
of the replication system versus the size of the network: that is, we
maintain the network density constant but we consider a number of
nodes ranging between 100 and 1000. As one could expect, the number of
replicas grows linearly with the network size, in
Fig.~\ref{fig:rep_network_size}, while
Fig.~\ref{fig:load_network_size} and Fig.~\ref{fig:delay_network_size}
show that the network size has virtually no impact on the average
workload at  replica nodes and on the delay, respectively. 

Overall, our replication scheme shows excellent scalability properties,
since it can dynamically adapt the number and placement of replicas to
the network settings, so as to maintain a constant utilization of communication
and memory resources at each node. Moreover, we recall that such result
is obtained with local measurements only, and thus the cost of the process
does not change with the number of items or the size and density of the
network.


\section{Benchmarking our replication scheme to other
  approaches \label{sec:other-approaches}} 

We now turn our attention to a network system where information
items are associated to different query rates, and we evaluate the
allocation of replicas for each content. 
In this case, we compare the performance of our replication scheme 
with that of the so-called square-root replication
strategy~\cite{sigcomm02}.  
According to such a strategy, the allocation percentage $\alpha(i)$
for a content $i$ is proportional to the square root of the total
demand per second for that content,
 i.e., 
\[ \alpha(i) = \frac {\sqrt{p(i)}} {\sum_{i=1}^{I} \sqrt{p(i)}}. \] 
In~\cite{sigcomm02}, it has been proved that square-root replication  
is optimal in terms of number of solved queries. Although initially  
introduced for wired, unstructured, peer-to-peer networks, the
square-root rule has since been  applied  to wireless
networks as well~\cite{icdcs09}.  

\begin{figure}[t]
\begin{center}
  \includegraphics[scale=0.3]{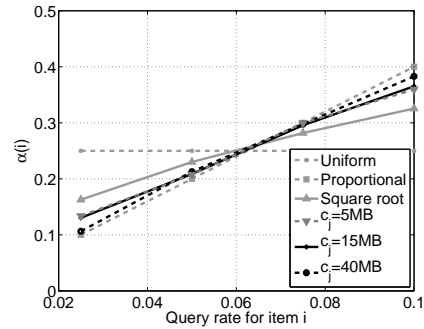}
\end{center}
\caption{Fraction of replicas for each of the four items, in comparison with
  uniform, proportional and square-root allocation\label{opt_rep_sr}}
\vspace{-3mm}
\end{figure}

We derive our simulation results in the case of $I=4$ items with
different popularity, and $c_j$=$\{5,15,40\}$~Mbytes.
Fig.~\ref{opt_rep_sr} shows the fraction of the total number of
replicas of item $i$, versus the associated query rate
$p(i) V\lambda$.  The plot compares our scheme with: (i) the square-root
strategy, (ii) a uniform strategy, which allocates the same number of
replicas per item, and (iii) a proportional strategy, where the number
of replicas is proportional to the content popularity.  We observe
that our scheme achieves an allocation in between the square-root and
proportional distributions, while it is far from that obtained under
the uniform strategy. This suggests that 
our replication scheme well approximates the
optimal replication strategy. In particular, we can observe that, when
$c_j$ is higher, i.e., replica nodes are more generous in reserving
resources to serve requests, the allocation tends to follow a
proportional distribution.  Conversely, in presence of lower values of
$c_j$, i.e., when the budget is limited, the allocation better
fits the square root rule.  In other words, a ``strict'' budget
sacrifices content replicas that play a marginal role in achieving 
low access cost: such replicas are dropped and the overall
shape of the distribution drifts from proportional to square root.

Before we move on, a further observation is required. Since our
replication scheme roughly achieves the result obtained by a
square-root allocation, 
it is reasonable to wonder why a different approach to
content replication is required. First of all, in this work we have
different objectives than that of \cite{sigcomm02}: load-balancing,
for example, requires an additional layer to complement the square
root allocation scheme, which instead we achieve as part of our
design. Furthermore, the distributed version of the replication
algorithms proposed in \cite{sigcomm02} has some limitations that
renders them less suitable to be deployed in a mobile, wireless
environment. The simple path replication scheme catering to low storage requirements, 
just like our scheme, substantially
over/undershoots the optimal number of replicas. The other approaches
discussed in \cite{sigcomm02} are better at converging to an optimal
number of replicas but require the bookkeeping of large amounts of
information. Finally, the design and the evaluation of such algorithms
in \cite{sigcomm02} do not take into account the dynamic nature that
is typical of a mobile network. 

As a second step in our comparative evaluation, we benchmark 
our replication mechanism  with  a
simple caching scheme. In particular, we consider a
\textit{pull-based}\footnote{It is not the focus of this work to explore
  push-based mechanisms, nor more advanced approaches such as
  interleaving of push/pull phases.
} caching mechanism: a node issues 
a query for an information item of interest to other nodes in its
vicinity. Such a request can travel up to $h$ hops away 
from the node that issued the request. If a request is not satisfied within a
timeout, the content is fetched directly from the cellular network.  
After having successfully downloaded the content, the node stores it
until the corresponding validity time expires. In case a node receives
a query for the stored content, it will serve it through
device-to-device communication. 
Note that, if a node is not interested in an information item, it will
not participate to the caching process, including content transfer and
storage. 

In summary, with the mechanism outlined above, information items spread
from one node to another in the network in a manner that loosely
resembles  an epidemic diffusion process. However, when this content
propagation is hindered by availability problems, the cellular
network is used to create new content sources and avoid starvation. 

With respect to the replication scheme we propose, the pull-based
caching approach analyzed here differs in many aspects. First,
such a caching scheme eventually achieves full content replication, in
that all nodes, at the end of the diffusion process, hold a copy of
the content and can serve requests from neighbors. Instead, the goal
of our replication mechanism is to find the optimal number of replicas
that minimize content access costs, while guaranteeing load balancing.
Additionally, in the caching scheme, nodes simply discard expired
content, while, in our scheme, replica nodes are in charge of downloading
up-to-date versions of the content.
Since in our simulations nodes are loosely synchronized, the former
behavior implies that, at regular intervals corresponding to the content 
version expiration times, the whole content diffusion process restarts from
scratch. 

In order to better understand our results, we now proceed with some 
key intuitions that follow from the
differences between caching and replication schemes outlined above.
It is well known that pull-based caching approaches are
sub-optimal during the bootstrap phase of the content delivery
process: the few nodes storing a copy of the content are overwhelmed
by queries originating from nearby nodes, while the vast majority of
the other nodes remain idle and wait for the content to propagate
towards them. The caching scheme we evaluate here partially overcomes
this problem by allowing nodes to fetch content through the cellular
network. However, it is reasonable to expect a
large number of ``external'' data transfers: as a consequence, access
congestion may arise also at the cellular level. 
Finally, we note that when the content is unpopular, the diffusion
process is even slower and the above negative effects are amplified. 

In the following, we test the performance of the replication and
caching approaches in presence of two content discovery mechanism: the
one presented in Sec.~\ref{sec:simulation} and employed in the
previous sections, which is based on a content location service, 
and a flooding-based approach.
The latter mechanism lacks the knowledge of replica node identities,
and thus floods the network with queries for the desired content,
although the overhead is reduced by means of a PGB-based, TTL-bounded
forwarding. The presence of two discovery techniques allows us to
comment on the impact that an optimized, yet complex solution (as the one
based on the use of a content location service) and 
a simple, yet sub-optimal one (flooding) have on the overall system
performance.

\begin{figure}[t]
\centering    
\subfigure[Number of replicas]{
	\label{fig:caching_replicas}
	\includegraphics[width=0.22\textwidth]{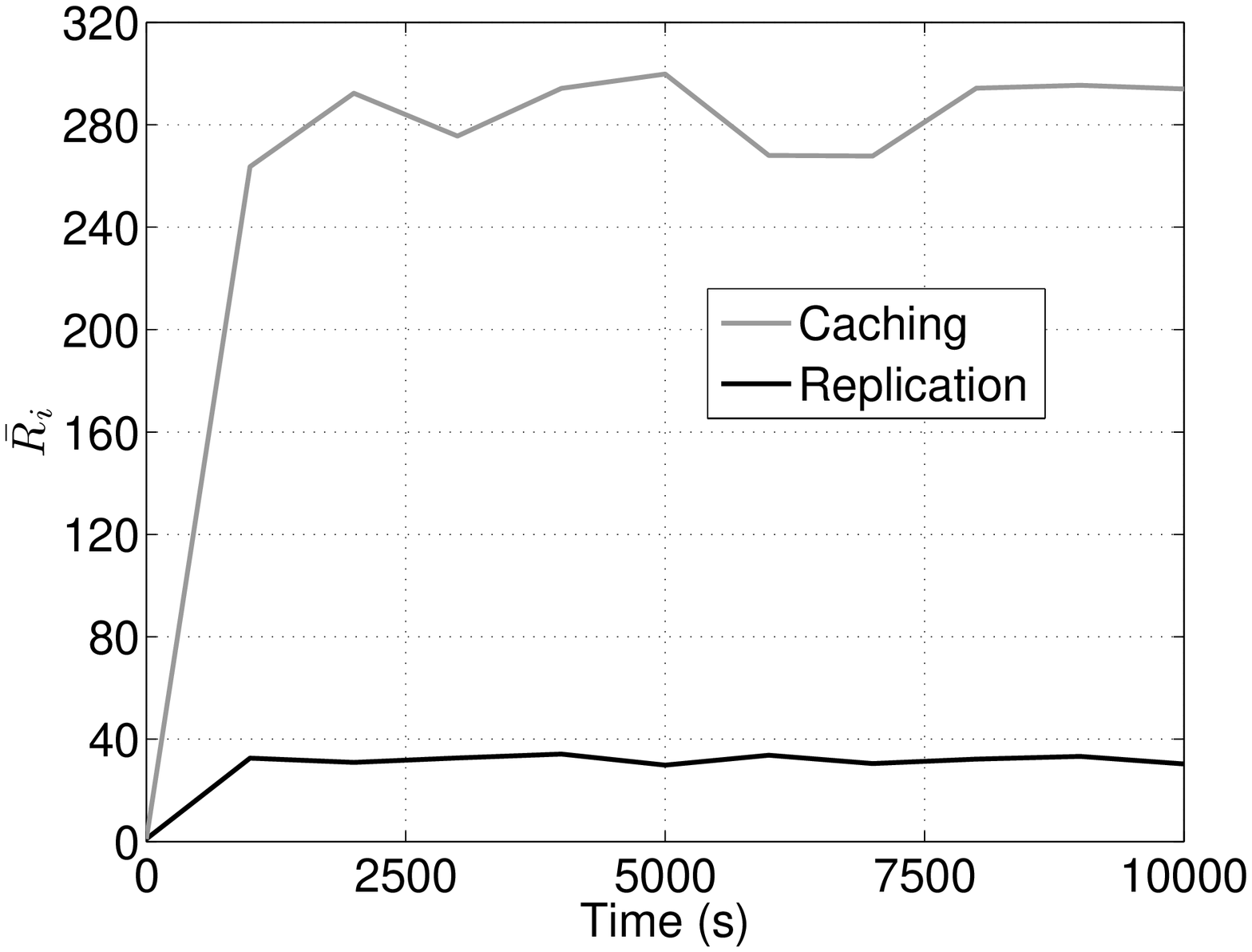}
}
\subfigure[$\chi^2$ index]{
	\label{fig:caching_chisquare}
	\includegraphics[width=0.22\textwidth]{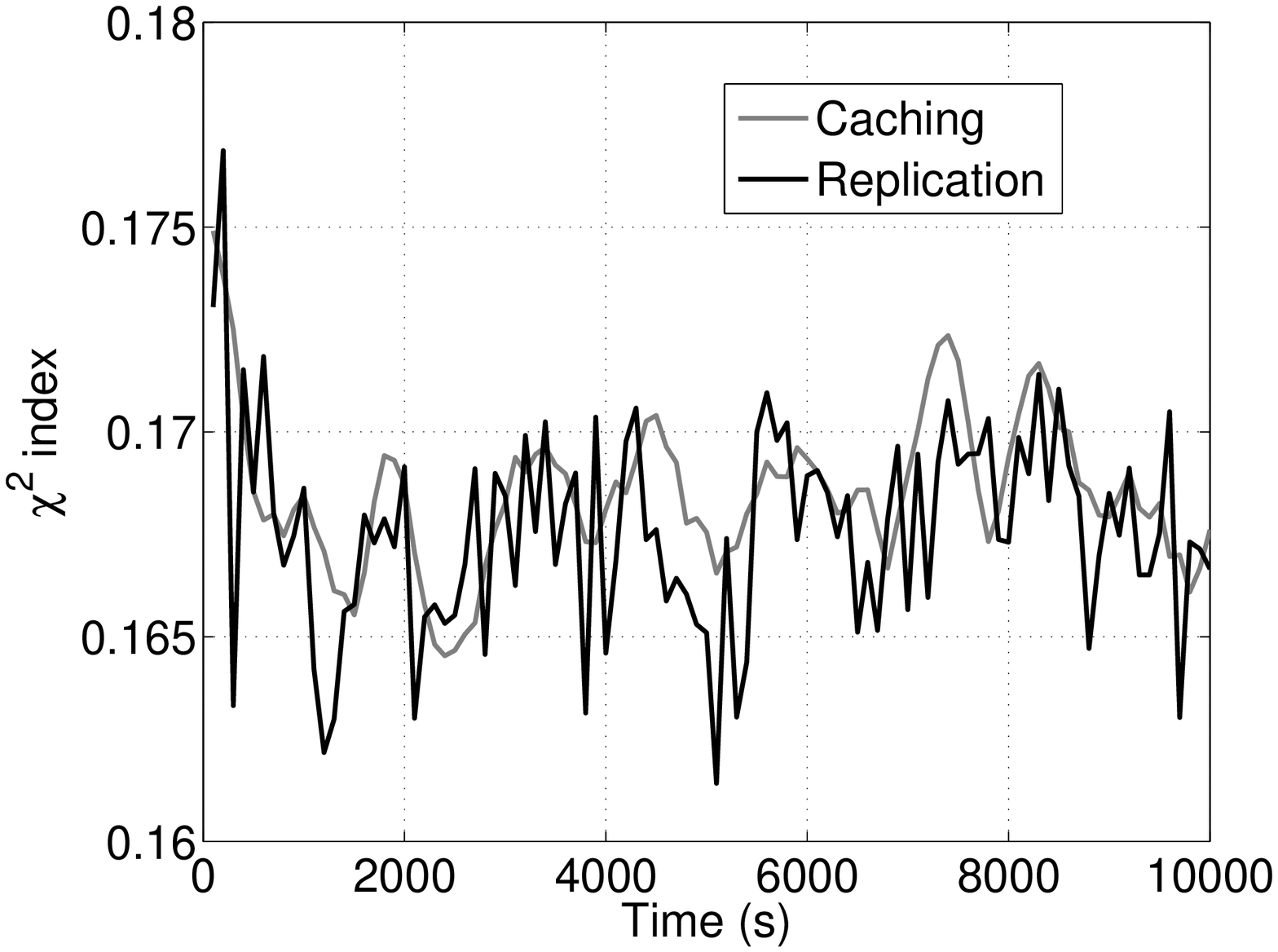}
}
\caption{Performance of caching and replication mechanisms in terms of
  (a) number of replicas and (b) $\chi^2$ index, for 100\% content popularity
  and 100~s content validity time}
\label{fig:performance_rep_cache}
\vspace{-3mm}
\end{figure}

We first focus on the behavior of the replication and caching schemes
over time. We run the two solutions in the identical standard settings
outlined in Sec.~\ref{sec:simulation}, assuming a content validity time
of 100~s and injecting one replica in the network at the beginning of
the simulation. The number of replicas present in the system over time
is depicted in Fig.~\ref{fig:caching_replicas}. We observe that,
while our replication scheme controls the number of replica nodes and
keeps it relatively small, the caching solution leads to a rapid growth
of users caching the content. As expected, by achieving full replication,
the caching strategy is more expensive than the replication scheme for
the mobile nodes, in terms of storage requirements.  

One may argue that fewer content replicas may lead to a suboptimal
placement: full replication ensures that the content resides where the
demand is. The results illustrated in
Fig.~\ref{fig:caching_chisquare}, however, show that such additional
storage space usage does not lead to any significant advantage in
terms of the quality of replica placement. The $\chi^2$ index
obtained by comparing the geographical distribution of replicas under
the two schemes with that computed by the centralized 
solution is essentially equivalent.


\begin{figure}[t]
\centering    
\subfigure[Query solving delay]{
	\label{fig:performance_delay}
	\includegraphics[width=0.22\textwidth]{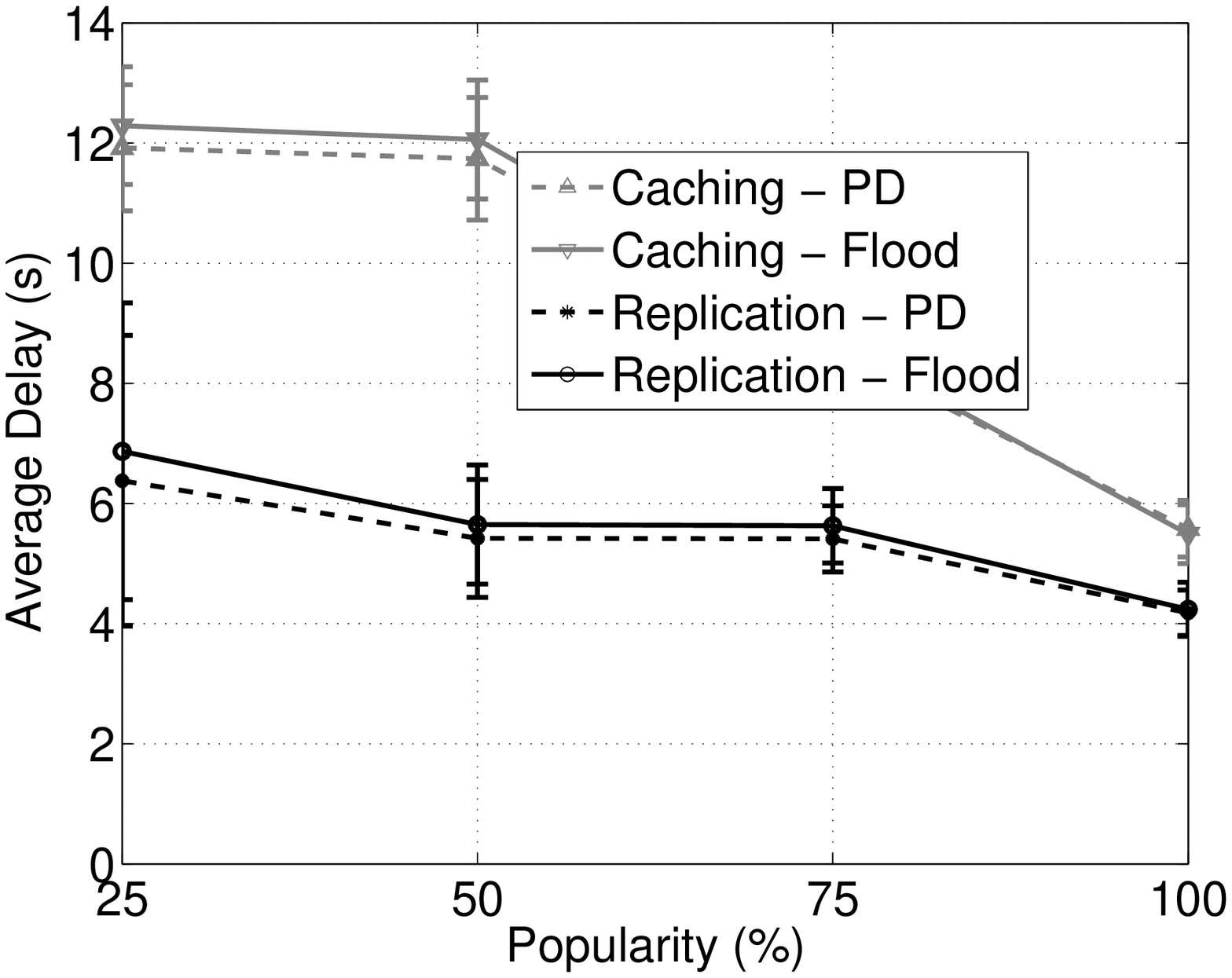}
}
\subfigure[Percentage of external downloads]{
	\label{fig:performance_externaldownload}
	\includegraphics[width=0.22\textwidth]{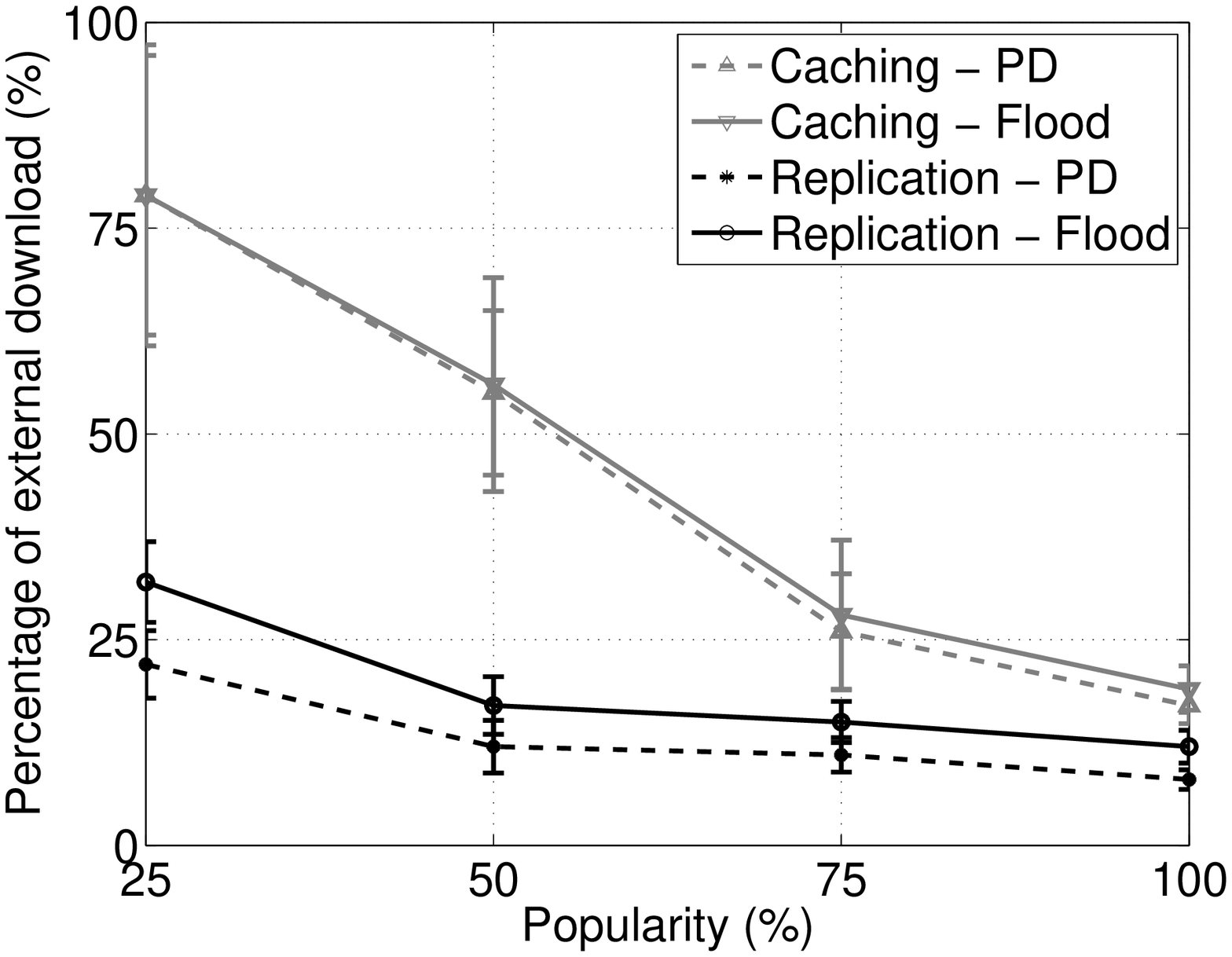}
}
\caption{Performance of caching and replication mechanisms in terms of
  ration of cellular downloads (a) and query solving delay (b) for content popularity of 25-100\%}
\label{fig:performance_rep_cache_pop}
\vspace{-3mm}
\end{figure}

\begin{figure}[t]
\centering    
\subfigure[Query solving delay]{
	\label{fig:performance_delay_updateperiod}
	\includegraphics[width=0.22\textwidth]{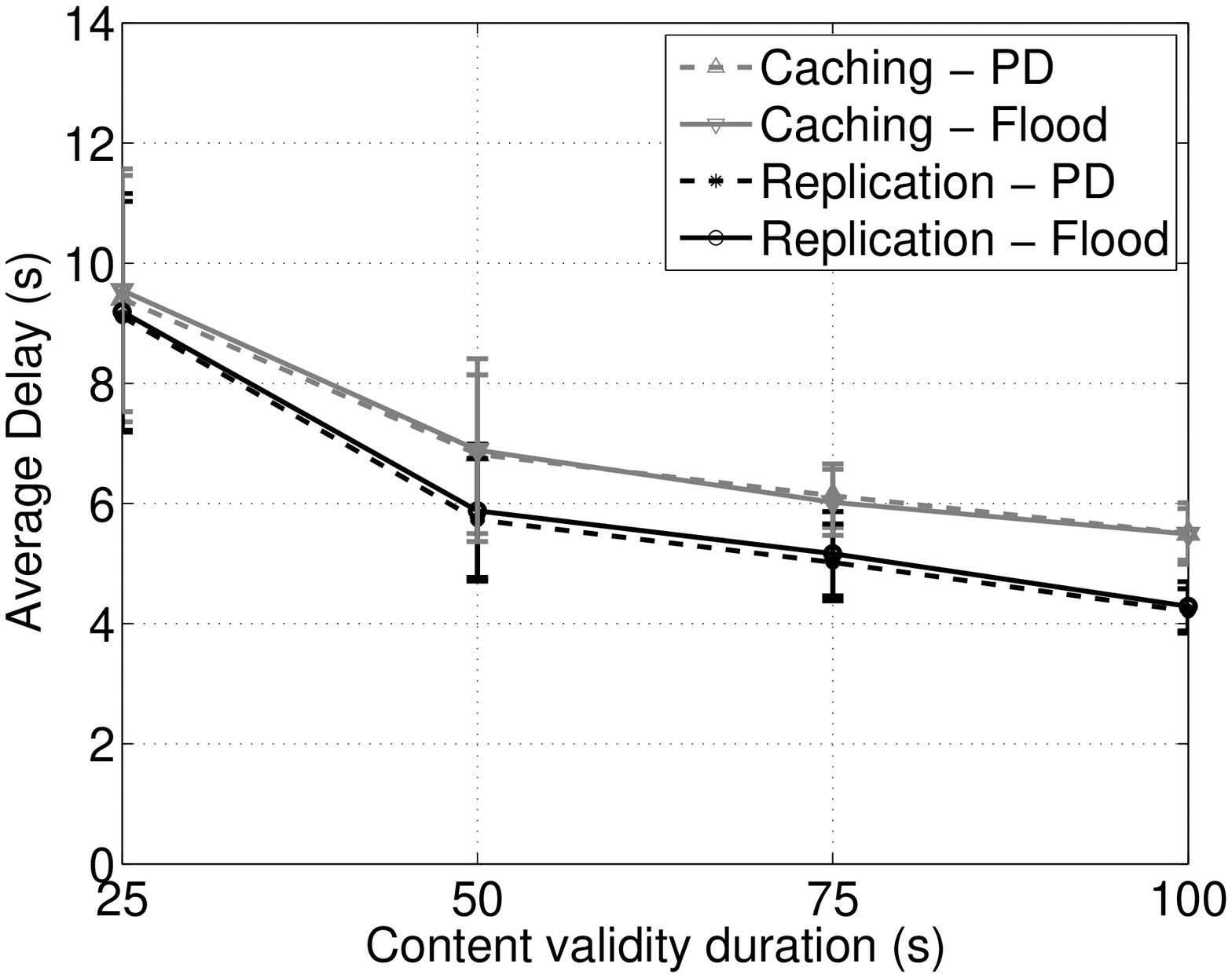}
}
\subfigure[Percentage of external downloads]{
	\label{fig:performance_externaldownload_updateperiod}
	\includegraphics[width=0.22\textwidth]{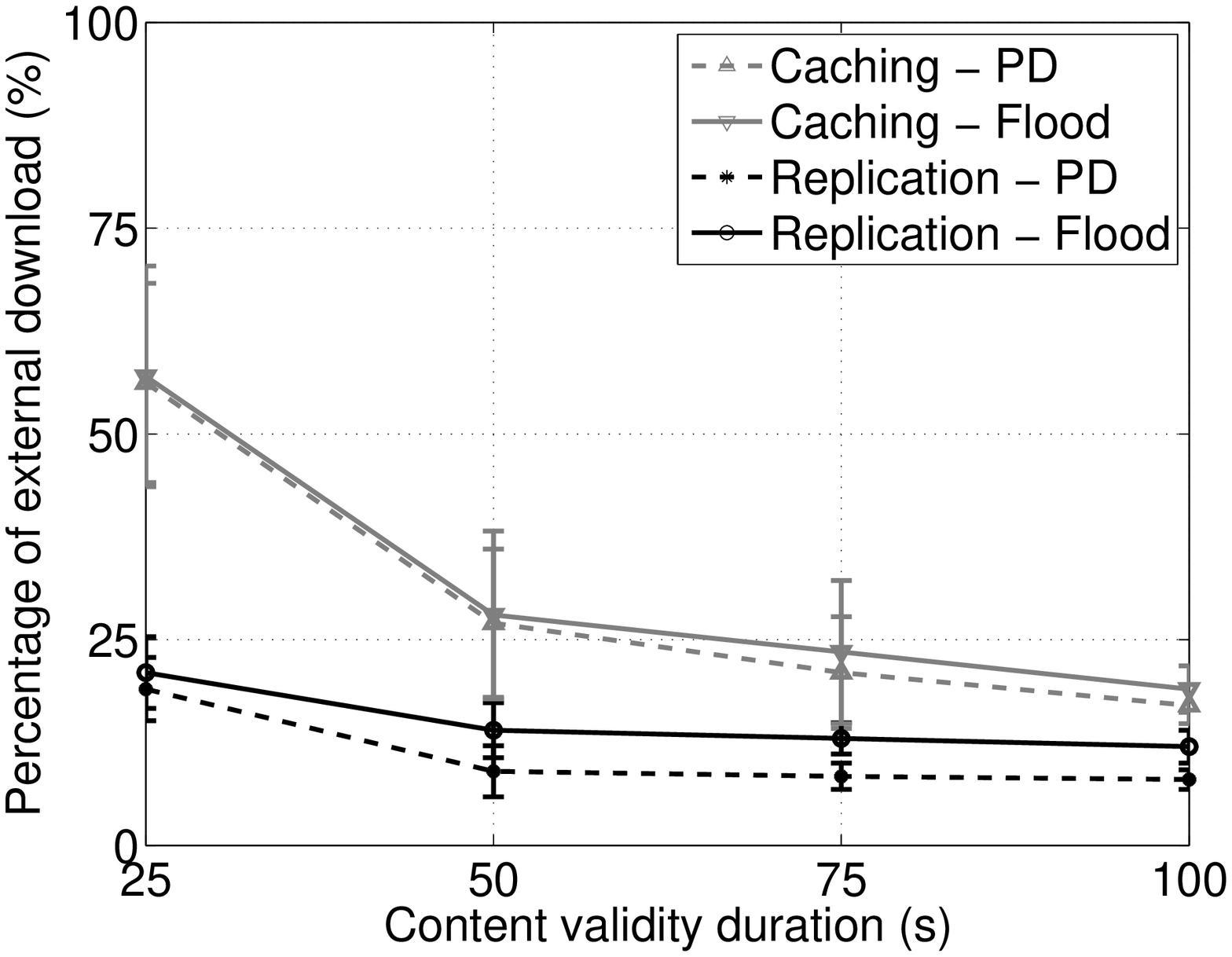}
}
\caption{Performance of caching and replication mechanisms in terms of
  ratio of cellular downloads (a) and query solving delay (b) for
  different content validity periods [25,50,75,100]~s}
\label{fig:performance_rep_cache_updateperiod}
\vspace{-3mm}
\end{figure}

We now compare the performance of the caching approach with that
of our replication scheme, when considering the following metrics 
that complement those previously employed:
\begin{itemize}
\item query solving delay, intended as the time elapsed from
the instant when a node sends the first query until  the
request is fulfilled, by either a replica node or the cellular
network;
\item percentage of external downloads, i.e., queries that resulted
in an external download, with respect to the overall requests generated
in the network.
\end{itemize}

Assume the content update period to be fixed at
100~s. Fig.~\ref{fig:performance_delay} shows the average delay (along
with the 95\% confidence interval)  
for the replication and caching scheme as
the content popularity varies.
As hinted at above, the replication scheme outperforms the caching
mechanism, and the difference in the relative performance is amplified
(in favor of replication), as the content popularity decreases. Indeed,
as content popularity decreases, fewer nodes participate in the
diffusion process that underlies the caching scheme. As such, nodes
have to wait longer for their queries to be satisfied and, in
general, they end up downloading the content from the cellular
network. Instead, when the content popularity is high, the
epidemic-style diffusion process performs better, and the delay
decreases. 
Fig.~\ref{fig:performance_externaldownload} reinforces the key
intuitions we discussed in this section: when the content diffusion
process is hindered by content popularity, mobile nodes resort to the
cellular network to compensate for the delays of device-to-device
communication. Our replication scheme outperforms the caching
approach also in this aspect: by approximating optimal content
replication and placement, our mechanism reduces the content access
costs, in terms of congestion. Instead, the caching mechanism does not
alleviate access congestion: i)  nodes in the vicinity of a
content replica will ``collide'' to obtain the content through
device-to-device communication, and ii)  nodes resorting to
the cellular infrastructure because of query 
timeout expiration also compete
for bandwidth. These interwined aspects are exacerbated when the
content becomes stale: with our approach, few replica nodes take care
of the update process, while, with the caching scheme we study
here, the whole content diffusion process has to start over.

Next, we delve into the impact of the content update frequency, and
compare the replication and caching scheme when the content validity
time is in the interval $\left[25,100\right]$~s. Here the content popularity is
set to 100\%.
Fig.~\ref{fig:performance_delay_updateperiod} shows the delay for
the replication and caching scheme as the update frequency decreases
(i.e., larger update times). When the update frequency is
high, both caching and replication suffer in terms of access
delay. Requests for an updated version of the content put under stress
the replication scheme, because few replica nodes are in charge of the
content update, and consumer nodes have to wait for the update process
to finish. Instead, as we argued above, the caching
scheme has to restart at every content update, and this is suboptimal.
Fig.~\ref{fig:performance_externaldownload_updateperiod} reinforces
the intuition that the caching scheme, in order to mitigate a slow
diffusion process, heavily relies on cellular communications, a
phenomenon that is exacerbated when the update frequency is
high. Instead, the replication scheme is essentially unaffected by the
update frequency with respect to the number of external downloads.

As described earlier in this section, we carried out our comparative
analysis using different content access mechanisms. As reported in our
results, there is no noticeable impact of using a simple flooding
technique versus a more sophisticated one based on content location
service. However, although we do not report the results here for sake
of conciseness, the workload payed by each node because of queries
being flooded in the network is larger than with an auxiliary service
helping nodes to target the closest replica. 

In light of the results discussed above, our content replication scheme
clearly emerges as a simple, efficient and performing alternative to
traditional mechanisms that distribute the content through
opportunistic communications among the nodes. By controlling the
number and the placement of content replicas, our
mechanism appears to be suitable especially when content popularity is
not 100\%, both for performance and cost-related reasons.

\section{Related work}\label{sec:related-work}

Simple, widely used techniques for replication are gossiping
and epidemic dissemination \cite{HaraEpidemic,Simplot}, where the information is forwarded to a randomly
selected subset of neighbors. Although our RWD scheme may resemble
this approach in that a replica node hands over the content to a
randomly chosen neighbor, the
mechanism we propose and the goals it achieves (i.e., approximation of optimal number of replicas)
are significantly different. 

Another viable approach to replication is represented
by quorum-based \cite{Hubaux} and cluster-based protocols \cite{Hassanein05}.
Both methods, although different, are based
on the maintenance of quorum systems or clusters,
which in mobile network are likely to cause an exceedingly
high overhead.
Node grouping is also exploited in \cite{Hara01,Hara03},
where groups of nodes with stable links are used
to cooperatively store contents and share information.
The schemes in \cite{Hara01,Hara03}, however, require
an a-priori knowledge of the query rate, which is 
assumed to be constant in time.
Note that, on the contrary,  our lightweight solution
can cope with a dynamic demand, whose estimate by the
replica nodes is used to trigger replication.
We  point out that achieving content diversity is  
the goal of \cite{Yin} too, where, however, cooperation is exploited 
among one-hop neighboring nodes only.

Threshold-based mechanisms for content replication
are proposed in \cite{Almeroth,Hara0457}. In particular,
in \cite{Almeroth} it is the original server that 
decides whether to replicate content or not, and 
where. In \cite{Hara0457},  nodes have limited storage
capabilities: if a node does not have enough free memory,
it will replace a previously received content with a new one, only
if it is going to access that piece of information more frequently
than its neighbors up to $h$-hops. Our scheme 
significantly differs from these works, since it is a totally distributed,
extremely lightweight mechanism, which accounts for 
the content demand by other nodes and ensures 
a replica density that autonomously adapts to the 
network dynamics.

Finally, relevant to our study are the numerous schemes proposed
for handling query/reply messages; examples are \cite{Chen},
which resembles the perfect-discovery mechanism, and 
 \cite{Estrin02,Vaidya} where queries are propagated
along trajectories so as to meet the requested information. 
Also, we point out that the RWD scheme
was first proposed in our work \cite{casetti09}.
That paper, however, besides being a preliminary
study, focused on
mechanisms for content handover only: no
replication or content access were addressed.

\section{Conclusions}\label{sec:conclusion}
We focused on content replication in mobile networks and we addressed
the joint problem of (i) establishing the number of content replicas
to deploy in the network, (ii) finding their most suitable location,
and (iii) letting users efficiently access content through
device-to-device communication.

We studied the above problems through the lenses of facility location
theory and proposed a distributed, lightweight scheme that builds on
(i) local search approximations of the multi-commodity capacitated
facility location problem and (ii) parallel random walk diffusion in
non-regular graphs.  We showed that, despite its simplicity and the
fact that it only leverages local measurements, our replication
solution can approximate with high accuracy the solution attained by
optimal centralized algorithms, while also guaranteeing a fair
balancing of the communication and memory resources demanded of nodes.
Additionally, the scheme we propose adapts to network dynamics, in
terms of content popularity, size and set cardinality, as well as user
number, density and mobility.

When compared to different approaches to content replication and
caching, our approach performs closely to square-root-based
replication, while it outperforms traditional caching techniques that
mimic an epidemic diffusion of the content, especially in the more
challenging settings of low content popularity and high frequency of
content updates.

\bibliographystyle{IEEEtran}
\bibliography{biblio}

\end{document}